\documentclass[11pt]{article}
\pdfoutput=1

\usepackage[english]{babel}
\usepackage{amsmath}
\usepackage{amssymb}
\usepackage{graphicx}
\usepackage{natbib}
\usepackage{appendix, bbm, color}
\usepackage{fullpage, multirow}
\usepackage{authblk}
\usepackage[pdftex]{hyperref}
\usepackage{booktabs}
\usepackage{amsthm}

\usepackage{bbm}
\usepackage{cancel} 
\usepackage{comment}

\usepackage[font={small,it}]{caption}

\theoremstyle{definition}
\newtheorem{definition}{Definition}
\newtheorem{proposition}{Proposition}

\def\balpha{\mbox{\boldmath $\alpha$}}
\def\bt{\mbox{\boldmath $t$}}

\providecommand{\keywords}[1]{\textbf{\textit{Keywords:}} #1}
\usepackage[normalem]{ulem}

\begin{document}
\title{Bayesian isotonic logistic regression via constrained splines: an application to estimating the serve advantage in professional tennis}
\author[1]{Silvia Montagna\thanks{silvia.montagna@unito.it}}
\author[2]{Vanessa Orani} 
\author[3]{Raffaele Argiento}
\affil[1]{ESOMAS Department, University of Turin, Turin 10134, Italy}
\affil[2]{IEIIT, National Research Council of Italy, Genoa 16149, Italy}
\affil[3]{Department of Statistical Sciences, Universit\`a Cattolica del Sacro Cuore, Milano}

\maketitle

\begin{abstract}
In professional tennis, it is often acknowledged that the server has an initial advantage. Indeed, the majority of points are won by the server, making the serve one of the most important elements in this sport. In this paper, we focus on the role of the serve advantage in winning a point as a function of the rally length. We propose a Bayesian isotonic logistic regression model for the probability of winning a point on serve. In particular, we decompose the logit of the probability of winning via a linear combination of B-splines basis functions, with athlete-specific basis function coefficients. Further, we ensure the serve advantage decreases with rally length by imposing constraints on the spline coefficients. We also consider the rally ability of each player, and study how the different types of court may impact on the player's rally ability. We apply our methodology to a Grand Slam singles matches dataset. 
\end{abstract}


\section{Introduction}

Predicting the outcome of tennis matches has attracted much attention within sport analytics over the years for a number of applications. For example, prediction models can provide coaches useful feedback about how players are improving over time and who they should be able to beat. Further, prediction models could help assess fan engagement and determine who is the favourite player, by how much, and who is currently the best player. See, for example, \cite{Glickman, Klassen, Barnett,  Newton, Gilsdorf, Gomes, Smith, Irons, KOV2016} and references therein. \par
%For example, \cite{Barnett} use a point-based model to show how published player statistics can be combined to predict the serving statistics, \cite{Gomes} examine how rally length decreases as the match progresses because of the increasing fatigue of the players, and \cite{Smith} studies how the number of shots during a match varies across three playing surfaces.  \par
It is nowadays generally acknowledged that the service is one of the most important elements in tennis. Indeed, it has been observed that the serving player wins more points than the receiving player in elite tennis \citep{Lees}. With the advances in racquet technologies, most top male players can hit service speeds of over 200 Kph. \cite{KMR} point out that if the serving speed reaches the receiver's reacting threshold, it becomes virtually impossible for the receiving player to return the ball. In the extreme, a strong serve strategy that gets rarely broken reduces the competitiveness of the game, and this may result in a loss of spectator interest. For this reason, the International Tennis Federation (ITF) monitors the importance of the serve and can undertake measures, such as slowing surface speeds, to ensure the game's combativeness is not endangered.  \par 
While it is reasonable to assume that the serve advantage gets lost as the rally length increases, there are only a few contributions in the literature attempting to quantify the serve advantage and relate it to rally length via a statistical model. An early contribution is given by \cite{DB}, where the authors describe the advantage of serving in elite tennis by comparing points won by both the server and the receiver for a given rally length. They conclude that the serve advantage is lost after the 4th rally shot on men's first serve. Subsequently, \cite{KOV} proposes a Bayesian hierarchical model to estimate player-specific serve curves that also adjust for the opponent rally abilities. In particular, the author uses an exponential decay function to model the decline in serve advantage plus a random effect, representing the difference between the rally ability of the opponents.\par 
In this paper, we focus on the role of the serve advantage in winning a point as a function of the rally length. Our approach falls into the Bradley-Terry class of models \citep{BradlyTerry} and is built upon of \cite{KOV}. We propose a Bayesian isotonic logistic regression model by representing the logit of the probability of winning a point on serve, $f$, as a linear combination of B-splines basis functions, with athlete-specific basis function coefficients. We point out that while the term isotonic is used to denote regression models where monotonicity is imposed everywhere, in our application we may also want to accommodate for monotonicity only in a subinterval of the function domain. The smoothness of $f$ is controlled by the order of the B-splines, while their shape is controlled by the associated control polygon $\mathcal{C}$ \citep{deBoor}. In particular, to ensure the serve advantage is non-increasing with rally length we constrain the spline function $f$ to be non-increasing by controlling its control polygon. This essentially results in imposing a constraint on the coefficients of the spline function. Further, we allow for the probability to win on serve to also depend on the rally abilities of the opponents. We note that the rally advantage component of the model draws on \cite{KOV}, but we extend it further to study how the different types of court (e.g., clay, hard) may impact on the player's rally ability. It is indeed well known that some players favour and perform better on particular surfaces (e.g., Nadal holds 11 French Open (clay) titles and 2 Wimbledon (grass) titles). Each surface material presents its own unique characteristics and provides different challenges to the players, with certain playing styles working better on some types of court and less effectively on others. For example, a grass court is the fastest type of court because of its low bounce capacity. Players must get to the ball more quickly than with clay or hard courts, thus players with stronger serve will generally perform better on grass. The rally advantage component of our model reflects how a player is likely to perform on a particular surface, and this in turns affects the win probability. \par 
Our contribution is twofold: first, the basis function decomposition allows for a more flexible modelling of the longitudinal curve for serve advantage than that attainable via an exponential decay function. Our hierarchical Bayesian framework further accommodates for the borrowing of information across the trajectories of the different athletes, and allows for out-of-sample prediction; second, our construction allows for the inclusion of covariates (e.g., the type of terrain) in the modelling of the rally abilities of the opponents. Therefore, it becomes possible to examine how covariates impact on the rally abilities, and ultimately on the distribution of the serve advantage curves. \par 
The remainder of the paper is organised as follows. In Section~\ref{data} we provide an overview of the data used for our analysis. In Section~\ref{model} we present our hierarchical Bayesian isotonic logistic model. In Section~\ref{comparison} we compare our model with \cite{KOV}. Section~\ref{results} presents the results of our real data analysis, and conclusions are outlined in Section~\ref{conclusions}.

\section{Grand slam data} \label{data}
We consider point-by-point data for main-draw singles Grand Slam matches from 2012 forward. Organisations such as the ITF and Grand Slam tournaments record some data on professional tennis matches, but rarely make it available to the public. In this paper, we use data scraped from the four Grand Slam websites shortly after each event by Jeff Sackmann\footnote{\label{Jeff}https://github.com/JeffSackmann}. The data is also available with the \texttt{R} package \texttt{deuce} \citep{deuce}. \par
There are four Grand Slam tournaments, namely, the Australian Open, French Open, Wimbledon, and US Open. These tournaments are subdivided in two types of associations: the Association of Tennis Professionals (ATP), containing all the matches played by male athletes, and the Women's Tennis Association (WTA), containing all matches played by female athletes. We consider the male and female tournaments separately, thus obtaining two datasets. For both datasets, we include players with three matches or more in the training data. For a more robust inference, we consider only rally lengths between zero and thirty, counting as zero the first shot played by the server. For both datasets, we extract the following variables: rally length, the series of return hits of the ball from a player to the opponent (an integer between zero and thirty); the ID (name and surname) of the players serving and receiving, respectively; an indicator variable denoting if the server wins the point; the tournament name, used to derive the type of court in the different tournaments. Indeed, the Australian Open and US Open tournaments are played over a hard court, the French Open is played on clay while Wimbledon is played on grass. Unfortunately, other information of potential interest, e.g. the serve's speed and direction, is not available for every rally. Table \ref{tab:1} reports some summary statistics about the dataset. In Table \ref{tab:rallies} we report the total number of rallies in the ATP and WTA tournaments, respectively. Short rallies, i.e. rally lengths smaller than or equal to $4$, constitute $90\%$ of the rallies played during the Grand Slam tournaments.

\begin{table*}\centering
	\begin{tabular}{@{}rrrrr@{}}\toprule
		& \multicolumn{4}{c}{\textbf{Tournament}} \\
		\cmidrule{1-5} 
		& \textbf{US Open} & \textbf{Aus Open} &\textbf{ French Open} & \textbf{Wimbledon} \\ \midrule
		\textbf{Matches}\\
				ATP & 147  & 106 & 192 & 214 \\
				WTA & 158  & 106 & 186 & 118 \\
		\textbf{Players}\\
				ATP & 108  & 107 & 132 & 107 \\
				WTA & 108  & 127 & 119 & 104 \\
		\bottomrule
	\end{tabular}
	\caption{Number of matches and players in the four Grand Slam tournaments by association from 2012 forward. The data was collected by Jeff Sackmann$^{\text{\ref{Jeff}}}$, and is also available with the \texttt{R} package \texttt{deuce} \citep{deuce}.}
	\label{tab:1}
\end{table*}

\begin{table*}\centering
	\begin{tabular}{@{}rrrr@{}}\toprule
		\cmidrule{1-4} 
		&\textbf{Short rallies} &\textbf{Long rallies} & \textbf{Total}\\ \midrule
		ATP & 130577 & 14933 & 145510\\
		WTA & 71592  & 10288 & 81880\\
		\bottomrule
	\end{tabular}
	\caption{Number of rallies by tournament. We define as short a rally whose length is less than or equal to 4. }
	\label{tab:rallies}
\end{table*}

\begin{figure}[!h]
\centering
	\begin{minipage}{7cm}
		\begin{center}
			\includegraphics[width=7cm,height=8cm]{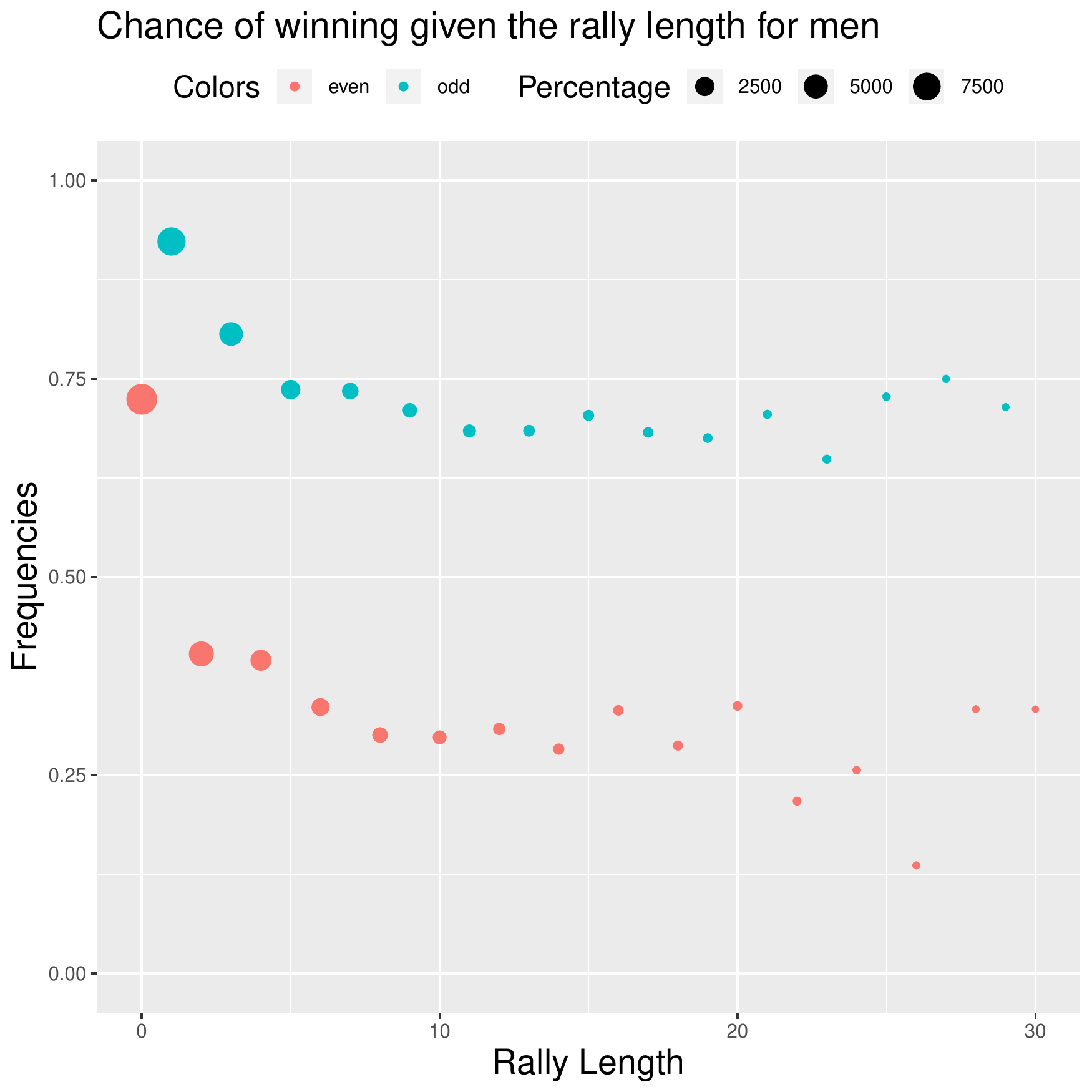}\\
		\end{center}
	\end{minipage}\ \
	\begin{minipage}{7cm}
		\begin{center}
			\includegraphics[width=7cm,height=8cm]{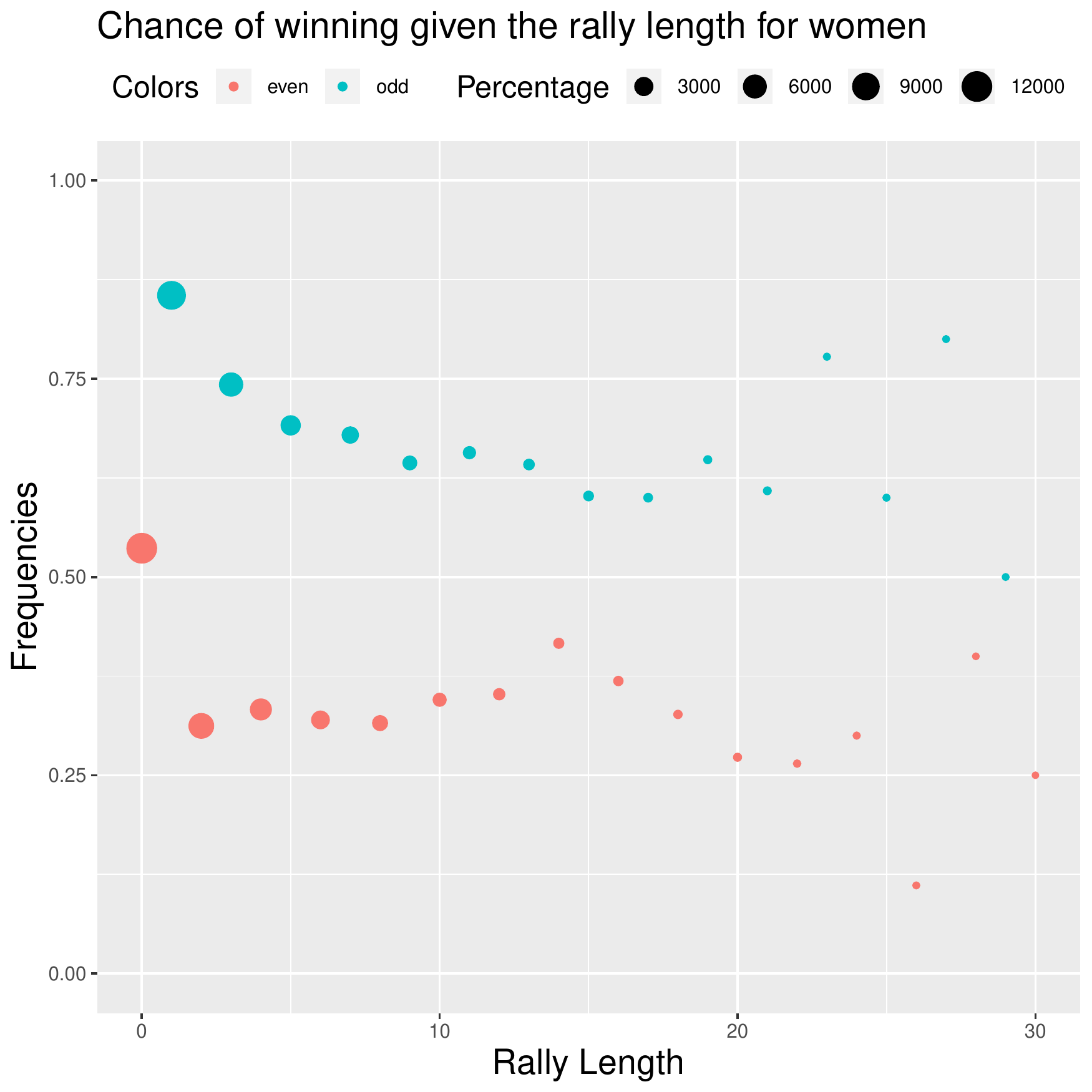}\\
		\end{center}
	\end{minipage}
	\caption{Observed frequency of rallies won by the server given the number of shots for the ATP tournament (left) and the for the WTA (right). The blue points represent the odd-shots, while pink points are the even-numbered rallies. The size of a point $(x,y)$ is proportional to the number of the server's victories with rally length equal to $x$ divided by the total number of points won by the server. }\label{fig:rally30}
\end{figure}

In Figure~\ref{fig:rally30} we report the observed relative frequency of rallies won by the server given the number of shots. It appears that the server has a higher chance of winning the point on odd-rally lengths compared to the even-rally lengths. This pattern can be explained by the fact that even-numbered rallies end on the server's racquet, so he/she can win or make a mistake. Since the $y$-axis report the observed relative frequency of winning for the server, all the even-shots in Figure \ref{fig:rally30} represent the case in which the server wins a point with a \textit{winner}. A \textit{winner} is a shot that is not reached by the opponent and wins the point. Occasionally, the term is also used to denote a serve that is reached but not returned into the court. On the other hand, the odd-numbered rallies are the winners or errors made by the receiver. In particular, the odd-shots in Figure \ref{fig:rally30} represent the errors done by the receivers. \par

\begin{figure}[!h]
\centering
	\begin{minipage}{7cm}
		\begin{center}
			\includegraphics[width=7cm,height=7cm]{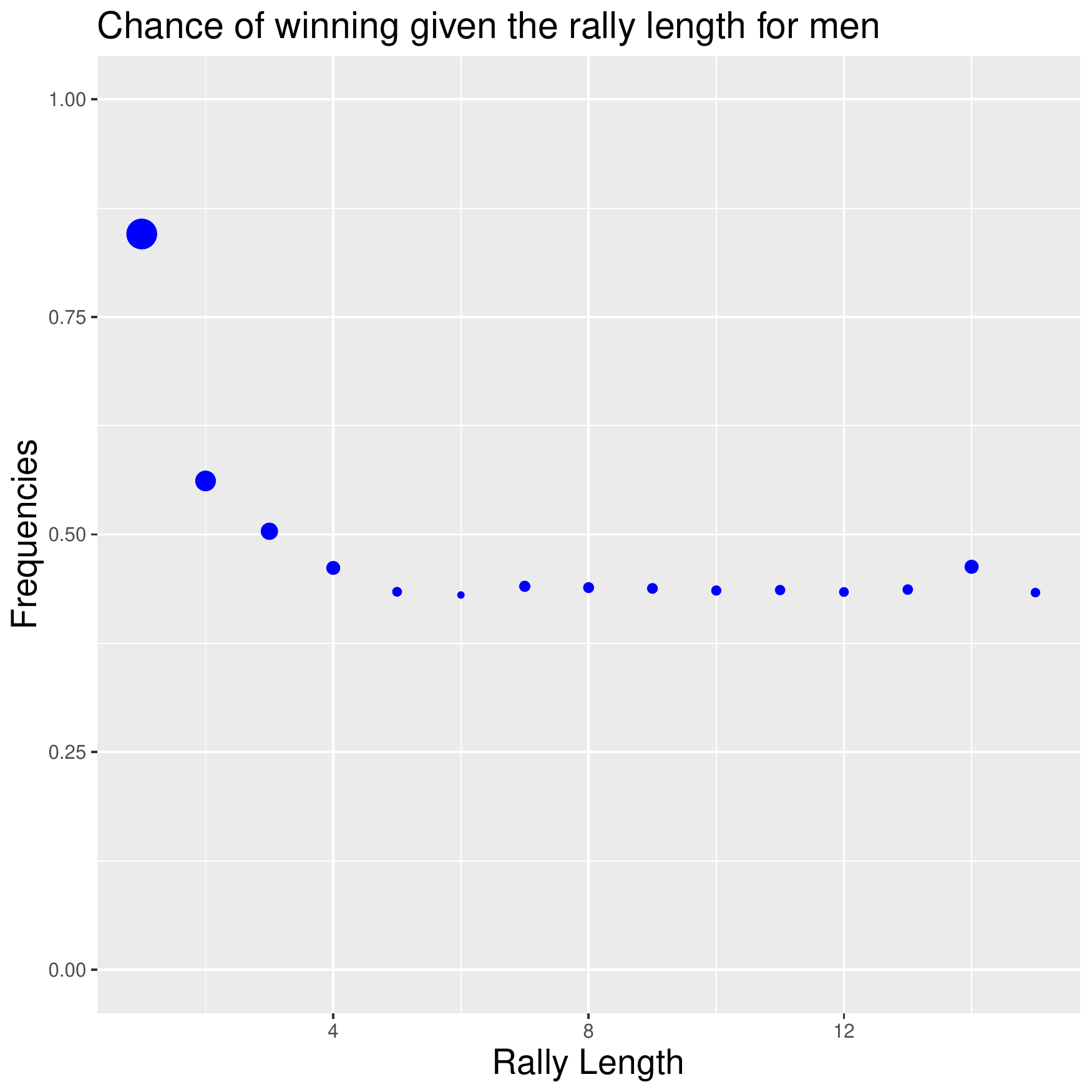}\\
		\end{center}
	\end{minipage}\ \
	\begin{minipage}{7cm}
		\begin{center}
			\includegraphics[width=7cm,height=7cm]{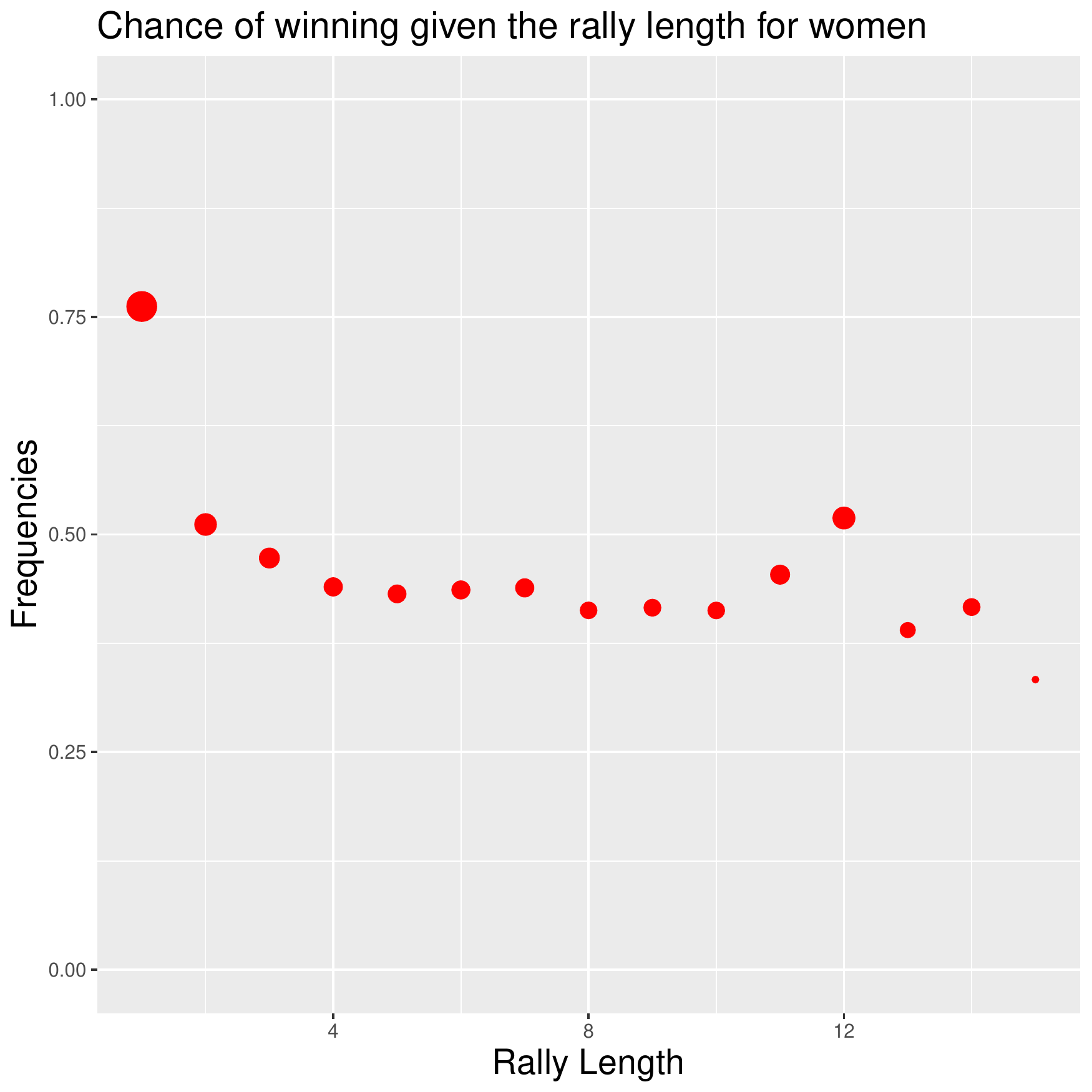}\\
		\end{center}
	\end{minipage}
	\caption{Conditional percentage of winning a point given the number of shots for the ATP tournament (left) and the for the WTA (right). Since we aggregated the odd and even results, rally length is between 1 and 15. The size of a point $(x,y)$ is proportional to the number of the server's victories with rally length equal to $x$ divided by the total number of points won by the server. \label{fig:rally15}}
\end{figure}

Because errors are more common than winners, we aggregate odd and even rally lengths (see also \cite{KOV}). We obtain a vector of integers, where 1 corresponds to rally lengths equal to zero or one, 2 corresponds to values 2 or 3 of rally length and so on. This ensures that the same set of outcome types for the server and receiver are represented within each group. The resulting frequencies are showed in Figure \ref{fig:rally15}. We observe that after the first shot the server's chance of winning the point drastically decreases. This is clear for both men and women. As conventional wisdom suggests, the server has the highest chance of winning a point at the beginning of the rally, owing to the strength of the serve. As the rally progresses, the serve advantage is expected to get increasingly small and have increasingly less influence on the outcome of the rally with each additional shot taken. 

\section{Hierarchical Bayesian isotonic logistic regression model}\label{model}

In this section, we present our Bayesian hierarchical isotonic regression approach to model the serve advantage. Let $x \in X $ be the discrete variable representing rally length, where $X = \{L,\ldots, U\}$ is the set of all integers between $L$ and $U$. Let $Y_{i,j}$ be a binary random variable which is equal to one if server $i$ wins the point against receiver $j$, and zero otherwise. We assume
\begin{equation} \label{eq:likelihood}
Y_{i,j}|p_{i,j}(x) \sim \text{Bernoulli}(p_{i,j}(x)), \quad x\in X
\end{equation}
where $p_{i,j}(x)$ is the probability that server $i$ wins a point against receiver $j$ at rally length $x$, e.g $\mathbb{P}[Y_{i,j}=1\vert x]$. We consider two components to model $p_{i,j}(x)$, the first describing the serve advantage and the second representing the rally ability of the players. Specifically, we model the logit of $p_{i,j}(x)$ as follows:
\begin{equation}
 \text{logit }\mathbb{P}(Y_{i,j}=1 \vert x) = \text{logit} \left[ p_{i,j}(x) \right] = f_i(x) + (\alpha_i -\alpha_j), \quad x \in X
 \label{eq:mainmodel}
\end{equation}
with
\begin{equation}
f_i(s) = \sum_{m=1}^{M}\beta_{i,m} b_{m}(s), \quad s \in [L, U] \subset \mathbb{R}
\label{eq:basis}
\end{equation}
for $i= 1,\dots, n_s,$ $j = 1,\dots, n_r$ and $i \neq j$, where $n_s$ is the total number of servers and $n_r$ the total number of receivers. The $\alpha$'s are athlete-specific parameters representing the rally ability of the player. We observe that the function $f(s)$ ($i$ index omitted for simplicity) is defined for each $s$ in the continuous interval $[L, U]$, however only its values $f(x)$, for $x$ in the discrete set $X \subseteq [L, U]$, enter the sampling model in Equation~\eqref{eq:mainmodel}. The continuous structure of $f(s)$ takes into account the overall trend of the serving advantage (i.e., it estimates the drop of the serving advantage as the rally length increases), accommodating for the longitudinal structure of the data. \par
The conditional log odds for the probability of the $i$-th server winning a point against the $j$-th receiver is a non-linear function of the rally length, $x$. Function $f_i(s)$ in Eq.~\eqref{eq:basis} represents the decay part of the model via a linear combination of basis functions $\{b_m(s)\}_{m=1}^M$, where $M$ is the dimension of the spline basis. In particular, we opted for B-splines basis functions of order $k$ on $[L, U] \subset \mathbb{R}$, and the $\beta_{i,m}$'s are athlete-specific basis function coefficients. Specifically, let $M \geq k \geq 1$ and $\bt \equiv \{t_m\}_{m = 1}^{M+k}$ be a non-decreasing sequence of knots such that $t_m < t_{m+k}$ for all $m$, and $t_k = L$ and $U = t_{M+1}$. Function $f_i(s)$ is a linear combination of the B-splines $b_1, \ldots, b_M$, and is called a spline function of order $k$ and knot sequence $\bt$ \citep{deBoor}. In other words, each $f_i$ is a piecewise polynomial of degree $(k-1)$ with breakpoints $t_m$, and the polynomials are $k - 1 - \mbox{Card}(t_m)$ times continuously differentiable at $t_m$. Here $\mbox{Card}(t_m)$ denotes the cardinality of $\{t_j: t_j = t_m\}$. Moreover, we recall that spline $b_m$ has support on the interval 
  $[t_m,t_{m+k}[$ and here we are going to assume $t_1=\dots=t_{k}=L$ and $t_{M+1}=\dots=t_{M+k}=U$.
  A more extensive presentation of spline functions is given in \cite{deBoor}. \par
In our model, the spline function is defined on the whole interval $[L, U]$ and does not go to zero for high values of rally length. Indeed, by looking at the last value of rally length in our application, e.g. $x = 15$, it is clear that the logit of the conditional probability of $i$ winning the point against $j$ reduces to
\begin{equation*}
\text{logit } \mathbb{P}(Y_{i,j}=1 \vert x=15) = \sum_{m=1}^{M}\beta_{i,m} b_m(15) + (\alpha_i -\alpha_j). \label{x15}
\end{equation*}
We consider this as an asymptote, describing the server's ($i$-th player) log-odds of winning a point against opponent $j$ when the serve advantage has vanished. We can interpret parameter $\alpha_i$ as the rally ability of the $i$-$th$ player. When the serve advantage vanishes, the probability of winning the point depends on the discrepancy between the rally abilities of the two players plus a constant, obtained from the B-splines basis.\par 
In the next sections, we provide more details regarding the modelling of the serve advantage and the rally ability, respectively. 

\subsection{Modelling the serve advantage}
\label{serveadvantage}
Hereafter we will denote a spline function as \emph{partially monotone} if $f_i(s)$ is monotone only in a sub-interval of its domain $[L,U]$, for example if $f_i(s)$ is monotone decreasing in $[L_0, U] \subset [L,U]$. To simplify the notation, we will omit index $i$ that denotes individual-specific objects, so we will let $f_i = f$ and $\beta_{i,m} = \beta_m$. Further, the spline coefficients $\{\beta_m\}_{m=1}^M$ will be also referred to as \emph{control points} in the following.

Given that the serve advantage is expected to decrease as rally length increases, the spline function $f(s)$ should be non-increasing in $[L, U]$. While the non-increasing behaviour can be directly learnt from the data for small values of rally length, this could be harder to achieve for large values of rally length due to data sparsity in this part of the function domain. In other words, we may have to impose that the spline function is non-increasing for large values of rally length.  Given a threshold $L_0$, we may allow $f(s)$ to be free to vary for small values of rally length (i.e., for $s<L_0$), while it is crucial to ensure that $f(s)$ is non-increasing as $s$ goes above  $L_0$. Thus, we would like $f$ to be partially monotone, according to our definition. Then, we need to investigate which condition the spline function must verify to guarantee the partial monotonicity constraint. To this end we will first provide the following definition.

\theoremstyle{definition}
\begin{definition}[\textbf{Control Polygon}] \label{def:controlpol}
  Let $\bt \equiv \{t_m\}_{m = 1}^{M+k}$ be a non-decreasing sequence of knots and let $f(s)=\sum_{m=1}^{M}\beta_mb_m(s)$ be a spline function of order $k > 1$ and knot sequence $\bt$. The \textit{control polygon} $\mathcal{C}(s)$ of $f(s)$ is defined as the piecewise linear function with vertices at $(\overline{t}_m, \beta_m)_{m = 1}^M$, where $$\overline{t}_m = \frac{t_{m+1} + \ldots + t_{m + k-1}}{k-1}$$ is called the $m$th knot average. 
  \end{definition}
  
\begin{figure}[h!]
\centering
\includegraphics[width=16cm, height = 8cm]{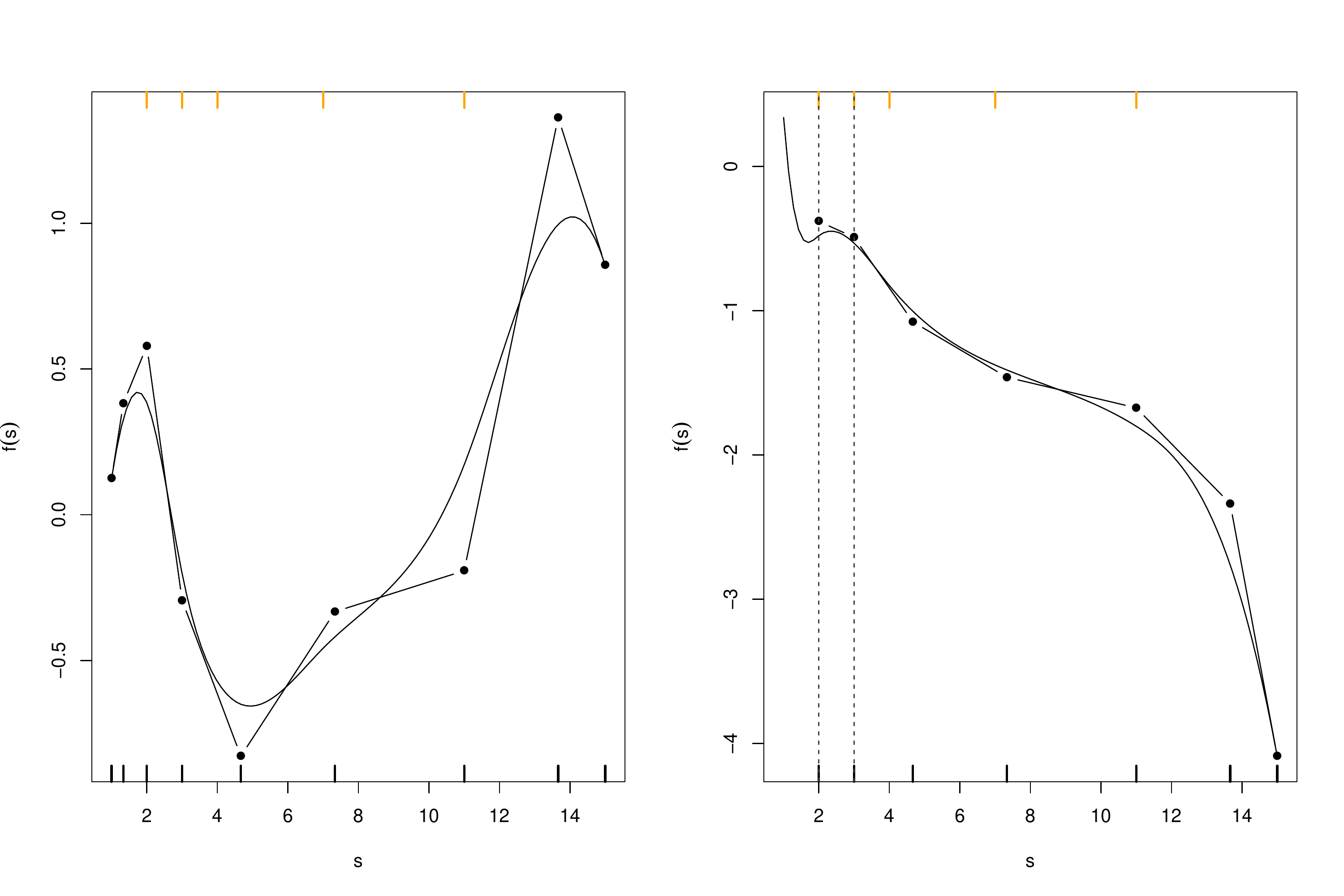}
\caption{Examples of spline functions (black solid lines) and associated control polygons (piecewise linear lines). In both panels, the spline functions were generated assuming B-splines functions of order $k = 4$ on $[1, 15]$ and with knot sequence $\bt = (1,  1,  1,  1,  2,  3,  4,  7, 11, 15, 15, 15, 15)$, and $M = 9$ is the dimension of the spline basis. Left panel: control points are generated as $\beta_m \stackrel{iid}{\sim}N(0,1)$, for $m = 1, \ldots, M$. Right panel: the control polygon is restricted to be non-increasing in $[\bar{t}_{m_{L_0} -k} = \bar{t}_3= 2, 15]$, and the resulting spline function is such from the smallest knot greater than $2$. Black rug bars indicate the knot averages, while orange rug bars denote the interior knots. }
\label{fig:splinefunction}
\end{figure}

We note that $\overline{t}_m < \overline{t}_{m+1}$ because it is assumed that $t_m < t_{m+k}$ for all $m$. The left panel of Figure \ref{fig:splinefunction} provides an illustrative example of a spline function with its associated control polygon. The spline function has order $k=4$ and knot vector $$\bt = (1,  1,  1,  1,  2,  3,  4,  7, 11, 15, 15, 15, 15)$$ The control points $\{\beta_m\}_{m=1}^M$ are randomly drawn from standard Normal distribution. The control polygon approximates the spline function $f$, and the approximation becomes more accurate as the number of control points increases. 

To ensure the serve advantage is non-increasing with rally length, we need to control the shape of the spline function $f$ on an interval $[L_0, U] \subseteq [L, U]$. To do so, we will follow the notation and the construction of \cite{france} hereafter. In particular, for all $s \in [L, U]$ we denote by $t_{m_s}$ the smallest knot greater than $s$, with the proviso that $m_s=M+1$ if $s$ belongs to $]t_{M},U]$. Further, we assume that $t_m < t_{m+1}$ for all 
$m \notin \{1,\dots,k-1,M+1,\dots,M+k\}$ so that $t_{m_s - 1} < s \leq t_{m_s}$ for all $s \in [L, U]$. We identify splines $b_{m_{L_0}}, \ldots, b_{M}$ as those whose support intersects with $[L_0, U]$. In other words, $[t_m , t_{m+k}] \cap [L_0, U] \neq \emptyset$ for $m \in \{m_{L_0}-k, \ldots, m_U-1\}$ and $[t_m , t_{m+k}] \cap [L_0, U] = \emptyset$ for $m \notin \{m_{L_0}-k, \ldots, m_U-1\}$, and the spline function restricted to the interval $[L_0,U]$
reduces to 
\begin{equation}
  f_{[L_0,U]}(s) = \sum_{m=m_{L_0} - k}^{M}\beta_{m} b_{m}(s), \quad s \in [L_0, U] 
\label{eq:basisreduced}
\end{equation}

  We define the restricted control polygon on $[L_0, U]$ as the control polygon
  associated to the spline \eqref{eq:basisreduced}, that is, the piecewise 
  linear function that interpolates the vertexes $P_m := (\bar{t}_m , \beta_m )$
  for $m = m_{L_0} - k,\dots, M$.  We denote by $\mathcal{C}_{[L_0,U]}(s)$ this
  restricted control polygon, which is defined for $s\in[\bar{t}_{m_{L_0-k}},\bar{t}_M]$. We also observe that $\mathcal{C}_{[L_0,U]}$ 
  is defined on a interval that contains $[L_0,U]$. Indeed 
  $\bar{t}_{m_{L_0-k}}\le L_0 $ and $\bar{t}_{M}=U$.

The spline function $f(s)$ can be restricted to be non-increasing on $[L_0, U]$ by imposing that the associated restricted control polygon $\mathcal{C}_{[L_0,U]}(s)$ is non-increasing as the following Proposition states.

\begin{proposition}
 Let $f(s)$ be a spline function with $s \in [L,U] \subset \mathbb{R}$, and let $\mathcal{C}(s)$ be the associated control polygon. Consider a real $L_0$ such that $L \leq L_0$. If the restricted $\mathcal{C}_{[L_0,U]}(s)$ is non-increasing on its support $[\bar{t}_{m_{L_0-k}},\bar{t}_M]$, then the restricted spline function $f_{[L_0,U]}(s)$ is also non-increasing on $[L_0,U]$. \label{prop:1}
\end{proposition}

The proof of Proposition \ref{prop:1} is given in Appendix \ref{A}. \cite{france} remark that if the control polygon is unimodal in $[L_0, U]$, then $f$ is unimodal or monotone on $[L_0, U]$, but it is possible to force $f$ to be unimodal by increasing the number of knots. 

Controlling the shape of the control polygon reduces to controlling the magnitude of the sequence of control points $\{\beta_m\}_{m=1}^M$. For example, for a non-increasing constraint on $[L_0, U]$, the broken line with vertexes $(\bar{t}_m, \beta_m)_{m_{L_0}-k}^M$ is non-increasing if $\beta_{m_{L_0}-k} \geq \cdots \geq \beta_{M}$. In particular, we impose that 
the spline coefficients of the restricted spline satisfy 
\begin{equation}
  \beta_{m}\le \beta_{m-1}\quad  m \in \left\{m_{L_0-k+1},\dots, M\right\}
  \label{eq:condizione_beta}
\end{equation}
%for $m \in \left\{m_{L_0-k},\dots, M\right\}$, coefficients $\{\beta_{m}\}_{m = m_{L_0}}^{M}$ are in the set $\Omega = \left\{ \bb \in \mathbb{R}^m | \beta_{m_{L_0}} \geq \cdots \geq \beta_{M} \right\}$ as proposed by \cite{france}.
Thus we have that the spline coefficients $\beta_1,\dots,\beta_{m_{L_0}-k}$ are free, while the spline coefficients $\beta_{m_{L_0}-k+1},\dots,\beta_M$ must be 
chosen such that condition in Equation~\eqref{eq:condizione_beta} is satisfied.
The right panel of Figure \ref{fig:splinefunction} shows an example of spline function whose shape is constrained to be non-increasing on $[L_0 = 3, U = 15]$. Given knots $\bt = (1,  1,  1,  1,  2,  3,  4,  7, 11, 15, 15, 15, 15)$, it is simple to realise that $m_{L_0} = 7$. Thus, $\beta_m \stackrel{iid}{\sim}N(0,1)$, for $m = 1, \ldots, m_{L_0 -k} = 3$, while the remaining coefficients are such that $\beta_3 \geq \beta_4 \geq \beta_5 \geq \ldots \geq \beta_{9}$. The resulting control polygon is decreasing from $\bar{t}_3 = 2$, and the spline function decreases from $L_0 = 3$. \par 

  \subsection{A new prior for the isotonic model} \label{sec:prior_isotonic}
  Now, let us return to the athlete-specific notation, that is, let's denote with $f_i(s)$ the spline function and with $\beta_{i,m}$ the spline coefficients for athlete $i$. Hereafter, we will construct the spline function $f_i(s)$ as (an intercept plus) a combination of $M$ B-spline functions defined on the closed set $[L,U] \subset \mathbb{R}$, with order $k$ and knot vector $\boldsymbol{t}=\left \{t_1,\dots,t_{M+k} \right\}$. For rally lengths $x \in [L,L_0]$, the decreasing behaviour of the spline function $f$ is  learnt from the data (Figure \ref{fig:rally15}), whereas the  non-increasing trend for rally length larger $L_0$ is assured by controlling the trend of to the restricted spline $f_{[L_0,U]}(s)$.

In order to specify a Bayesian model which takes into account the constraints on $f_i(s)$, we have to specify a prior distribution on the spline coefficients $\beta_1,\dots,\beta_M$ such that the conditions described in section~\ref{sec:prior_isotonic} are satisfied. 
With this goal in mind, the free spline coefficients are given a Normal prior distribution with mean $\beta_{m}$ and variance $\sigma^2_{\beta_m}$:

\begin{equation}
  \beta_{i m}|\beta_{m},\sigma_{\beta_m }^2 \overset{\text{iid}}{\sim} \mathcal{N}(\beta_{m},\sigma_{\beta_m }^2) \qquad \text{for} \qquad m = 1,\cdots, m_{L_0}-k
\label{eq:priorbeta}
\end{equation}
The prior mean and precision $\tau^2_m= \frac{1}{\sigma_{\beta_m }^2}$ are given conditionally conjugate prior distributions:

\begin{equation}
\beta_{m}|\beta_{0},\sigma^{2}_{\beta_0} \sim \mathcal{N}(\beta_{0},\sigma^{2}_{\beta_0}) \quad \text{and} \quad \tau^2_m|r_{\tau},s_{\tau} \sim \Gamma\left(\frac{r_{\tau}}{s_{\tau}^2},\left(\frac{r_{\tau}}{s_{\tau}}\right)^2\right),
\end{equation}
where $r_{\tau}$ is the mean and $s_{\tau}$ is the variance of $\tau_m^2$. Further: 

$$\beta_{0} \sim \mathcal{N}(0,100), \qquad \frac{1}{\sigma_{\beta_0}^2}  \sim \Gamma(0.1,0.1)$$ 
$$r_{\tau} \sim \mathcal{U}(0,10), \qquad  s_{\tau} \sim \mathcal{U}(0,10)$$
With regard to the constrained coefficients, we need to ensure the condition in Equation~\eqref{eq:condizione_beta} is verified. Therefore, we define these parameters recursively by letting 
\begin{equation}
  \beta_{i, m} := \beta_{i, m-1} - \varepsilon_{i, m}, \quad m=m_{L_0}-k+1,\dots,M 
\label{eq:priorbeta2}
\end{equation}
where  $\varepsilon_{i,m}$ are  random decrements  with
\begin{equation}
\varepsilon_{i,m}|r_{\varepsilon},s_{\varepsilon} \sim \Gamma\left(\frac{r_{\varepsilon}}{s_{\varepsilon}^2},\left(\frac{r_{\varepsilon}}{s_{\varepsilon}}\right)^2\right), \qquad \text{where} \quad r_{\varepsilon} \sim \mathcal{U}(0,10) \quad \text{and} \quad s_{\varepsilon} \sim \mathcal{U}(0,10)
\label{eq:priorepsilon}
\end{equation}

The last equation shows how a spline basis can be easily constrained to be non-increasing, while still retaining its essential flexibility. \par
For our application, we choose the same setup discussed in the example of Section~\ref{serveadvantage}. The constrain $L_0=3$ satisfies the empirical conclusion of \cite{DB}, namely, that the serve advantage is lost after the 4th rally shot. We assume independent Normal priors as in Equation \ref{eq:priorbeta} for the first three spline ($m=1,\dots,3$) coefficients and we adopt the recursive construction \eqref{eq:priorbeta2}, with the prior in \eqref{eq:priorepsilon}, for the remaining ($m=4,\dots,9$) coefficients.

\subsection{Modelling the rally ability}
\label{sec:rallying}
As outlined above, parameter $\alpha_i$ in Equation~\eqref{eq:mainmodel} can be interpreted as the rally ability of server $i$. It is clear that parameters $\alpha_i$ and $\alpha_j$ in Equation~\eqref{eq:mainmodel} are not identifiable, that is, adding and subtracting a constant to these parameters leaves $(\alpha_i - \alpha_j)$, thus inference, unchanged. Non-identifiability of the $\balpha$'s is not a concern if one is solely interested in learning the logit of $p_{i,j}(x)$. However, we are also interested in direct inference of the rally ability parameters, thus we need to include an identifiability constraint \citep{gelfand1999identifiability,Baio}. In particular, we adopt a sum-to-zero constraint by setting
$$\sum_{i=1}^{N}{\alpha_i}=0,$$
where $N$ is the total number of players in the dataset. We specify a Gaussian prior distribution for the rally ability parameter as in \cite{KOV}:
\begin{equation}
\alpha_i|\alpha_{ 0},\sigma_{\alpha} \sim \mathcal{N}(\alpha_{0},\sigma^2_{\alpha}), \qquad {\text{for}} \qquad i = 1, \ldots, N-1 \label{modelalpha}
\end{equation}
Finally, we specify the following conditionally conjugate non-informative priors on the hyperparameters:
$$\alpha_0 \sim \mathcal{N}(0,100) \qquad \qquad \frac{1}{\sigma^2_{\alpha}} \sim \Gamma(0.1,0.1) $$

\subsection{Estimating a court effect} \label{sec:future}
In the Grand Slam tournaments, players play over three different types of court: clay, grass, and hard, respectively. The tennis season begins with hard courts, then moves to clay, grass, and back again to hard courts. While very popular in the past, nowadays only Wimbledon is played on grass. Each surface elicits different ball speed, bounce height, and sliding characteristics. For example, grass courts produce little friction with the ball, which will typically bounce low and at high speed on this court. Conversely, clay slows down the ball a little and allows players more time to return it, resulting in longer rallies. Players have to adapt their technique effectively to the surface. However, adapting training and playing schedules is extremely physically demanding on the player. As a result, it is very difficult for one player to dominate across all the courts, and thus all the slams \citep{Starbuck2016}. \par
It is therefore reasonable to state that the surface type can impact on a player's performance. \cite{atp} study the effect of the different courts for ATP players using a Bradly-Terry model and conclude that taking this information into account leads to improved rankings of the players. In our model, it is straightforward to include court as a covariate within a regression model for the rally ability of each player, and observe the best player for each court. In particular, we define the probability of winning a point on serve given both the rally length and the type of court:
\begin{equation}
\text{logit } p_{i,j}(x,c) = \text{logit }\mathbb{P}[Y_{i,j}=1\vert x,c] = \sum_{m=1}^{M}\beta_{i,m} b_{m}(x) + (\alpha_{i,c} -\alpha_{j,c} ),
\label{modelcourt}
\end{equation}
for $i= 1,\dots, n_s, j = 1,\dots, n_r$ and $i \neq j$, where $n_s$ is the total number of servers, $n_r$ the total number of receivers, and $M$ is the dimension of the splines basis. Here index $c$ denotes the type of court, with $c \in \left\{1,2,3\right\}$, where 1 means clay, 2 stands for grass and 3 means hard. \par
Adding this covariate to our model does not affect the serve advantage, which is modeled as in Section~\ref{serveadvantage}. Conversely, we now have a subject-specific vector of rally abilities $\balpha_i = (\alpha_{i,1}, \alpha_{i,2}, \alpha_{i,3})^\top$, where $\alpha_{i,c}$ refers to court type $c$. To ensure the $\balpha_i$'s are identifiable for all players, we impose 
 $$\sum_{i=1}^{N}\sum_{c=1}^3 \alpha_{i,c} = 0,$$
where $N$ is the total number of players in the dataset. We specify a Gaussian prior distribution on the rally ability parameters:
\begin{equation}
\alpha_{i,c}|\alpha_{ 0},\sigma_{\alpha} \sim \mathcal{N}(\alpha_{0},\sigma^2_{\alpha}), \label{alphacourt}
\end{equation}
for $i = 1, \ldots, n-1$, $c=1, 2, 3$, and for $i = n$ and $c = 1, 2$. Finally, we specify the following conditionally conjugate non-informative priors on the hyper-parameters:

$$\alpha_0 \sim \mathcal{N}(0,100) \qquad \qquad \frac{1}{\sigma^2_{\alpha}} \sim \Gamma(0.1,0.1) $$

\section{Model comparison}\label{comparison}

In this section, we compare different models in order to identify the best on the tennis data. In particular, we consider the pair-comparison exponential decay model, proposed by \cite{KOV}, and three versions of our Bayesian isotonic logistic regression (BILR) model: 1) no constraints on the spline coefficients, thus the spline function is free of monotonicity constraints; 2) set $L_0 = 3$ ($U=15$) and impose an order constraint on the coefficients of the B-splines with support in $(3, 15]$, thus the resulting spline function is non-increasing in $(3, 15]$ (partially monotone); and 3) spline function constrained to be non-increasing in $[1, 15]$, with an order constraint on all basis function coefficients, $\beta_1 \geq \beta_ 2 \geq \ldots \geq \beta_M$. Setting 2) draws on \cite{DB}, who observe the serve advantage is lost after the 4th rally shot on men's first serve.

To compare the performance of the different methods, we compute four goodness of fit indices broadly used in the Bayesian framework. In particular, we consider the Log Pseudo Marginal Likelihood (LPML) \citep{LPML}, which derives from predictive considerations and leads to pseudo Bayes factors for choosing among models. Further, we compute the Deviance Information Criterion (DIC) \citep{DIC}, which penalizes a model for its number of parameters, and the Watanabe Akaike information criterion (WAIC) \citep{WAIC}. The latter can be interpreted as a computationally convenient approximation to cross-validation and it is not effected by the dimension of the parameter vector. Finally, we also compute the root mean squared error (RMSE).  \par

Prior to implementing the BILR model, one has to choose the order of the B-spline bases $k$, the number of knots and their location, which together determine the dimension of the spline basis. We recall that $M$ in Eq.~\eqref{eq:basis} is determined as $M = k$  $+ $ number of interior knots. 
Further, one has to choose the sub-interval of the spline function domain where monotonicity is to be imposed. For setting 2) above, this sub-interval is chosen to be $(3, 15]$. We performed some preliminary sensitivity analysis to investigate changes in performance of the BILR model due to different choices for the number of bases, their degree, the knots location, and range $[L_0, U)$.  Results of our sensitivity analysis are reported in \cite{tesiOrani}. For all three versions of our BILR model 1)-3) above, the spline functions $f_i(s)$, with $i = 1, \ldots, n_s$, are constructed from a B-spline basis of dimension $M = 9$ as in Eq.~\eqref{eq:basis}, defined on the closed interval $\left[L,U\right]=\left[1,15\right]$. The spline functions have order $k=4$ and knot vector $\bt = (1,  1,  1,  1,  2,  3,  4,  7, 11, 15, 15, 15, 15)$. This model maximises the LPLM criterion and minimises the DIC and the WAIC, as reported in \cite{tesiOrani}. 

\begin{figure}[h!]
	\centering
	\begin{minipage}{7cm}
		\begin{center}
			\includegraphics[width=7cm,height=6cm]{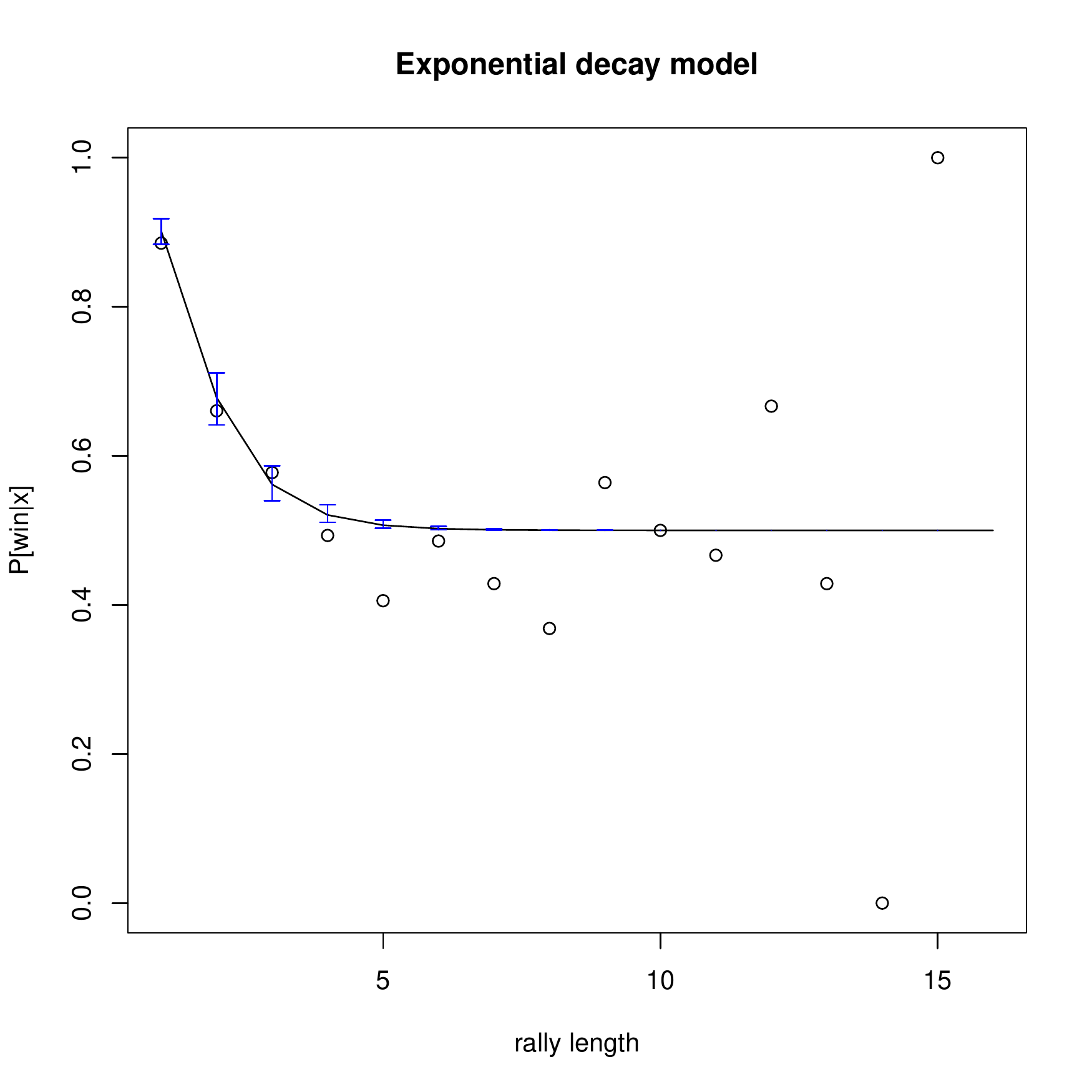}\\
		\end{center}
	\end{minipage}\ \
	\begin{minipage}{7cm}
		\begin{center}
			\includegraphics[width=7cm,height=6cm]{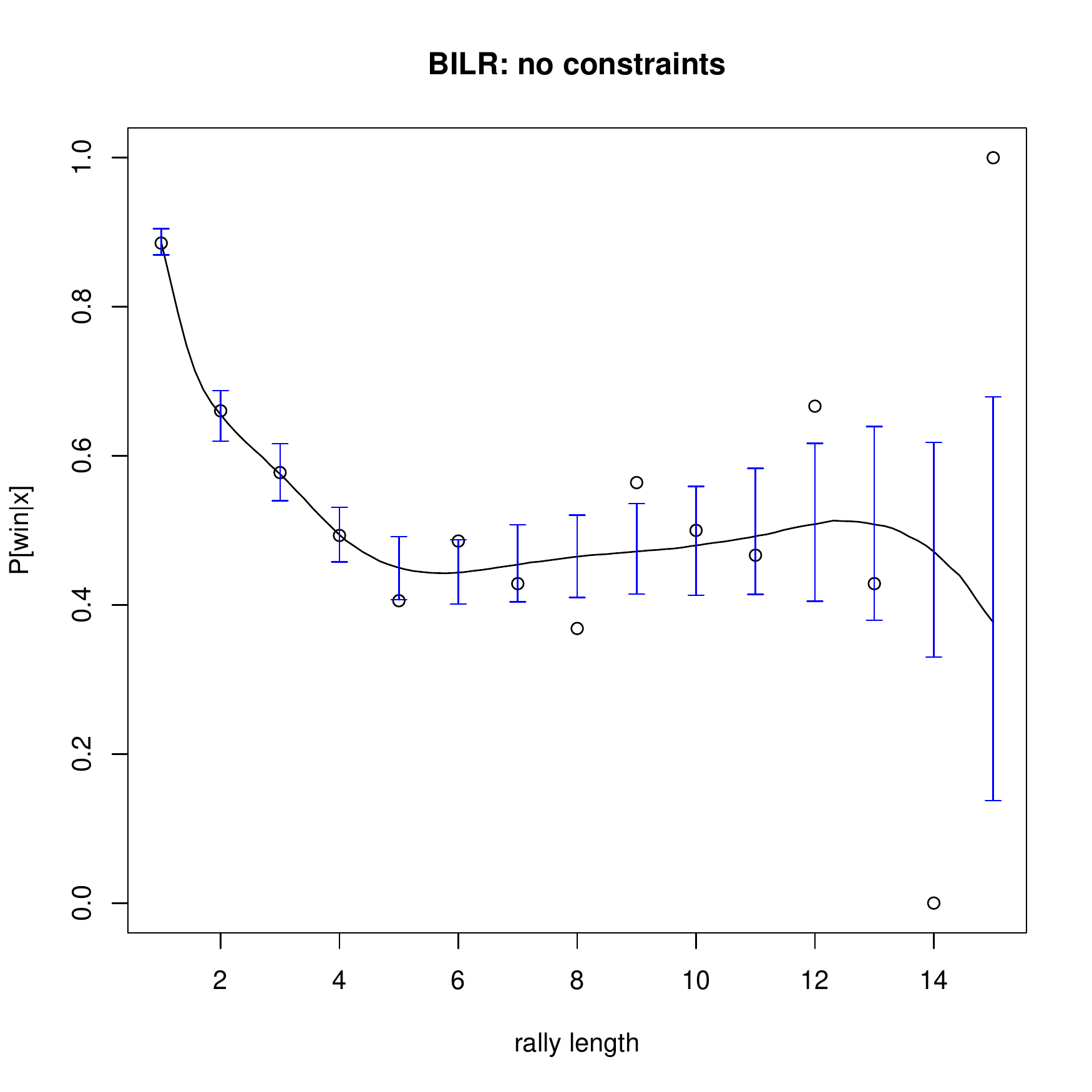}\\
		\end{center}
	\end{minipage}\ \
	\begin{minipage}{7cm}
		\begin{center}
			\includegraphics[width=7cm,height=6cm]{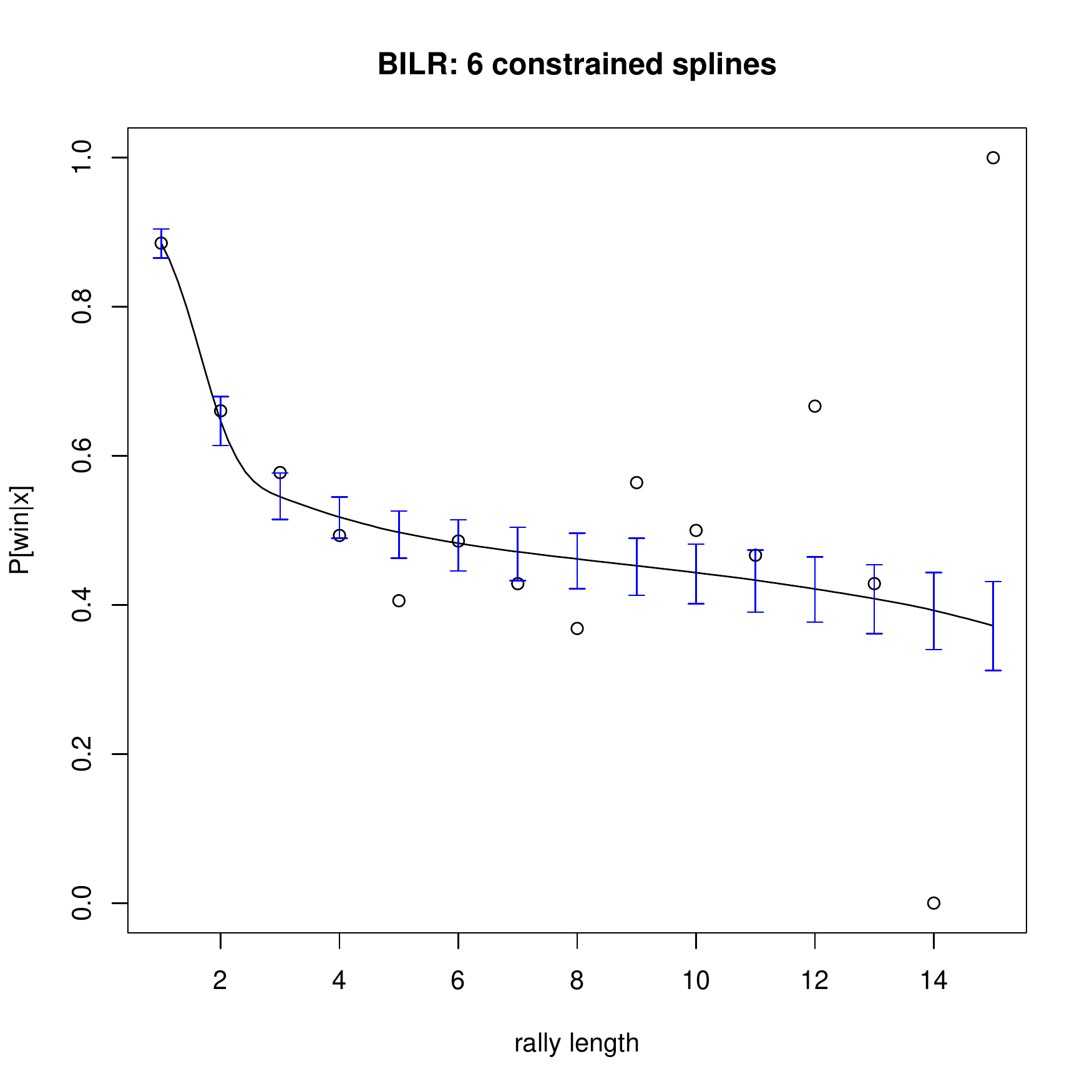}\\
		\end{center}
	\end{minipage}\ \
	\begin{minipage}{7cm}
		\begin{center}
			\includegraphics[width=7cm,height=6cm]{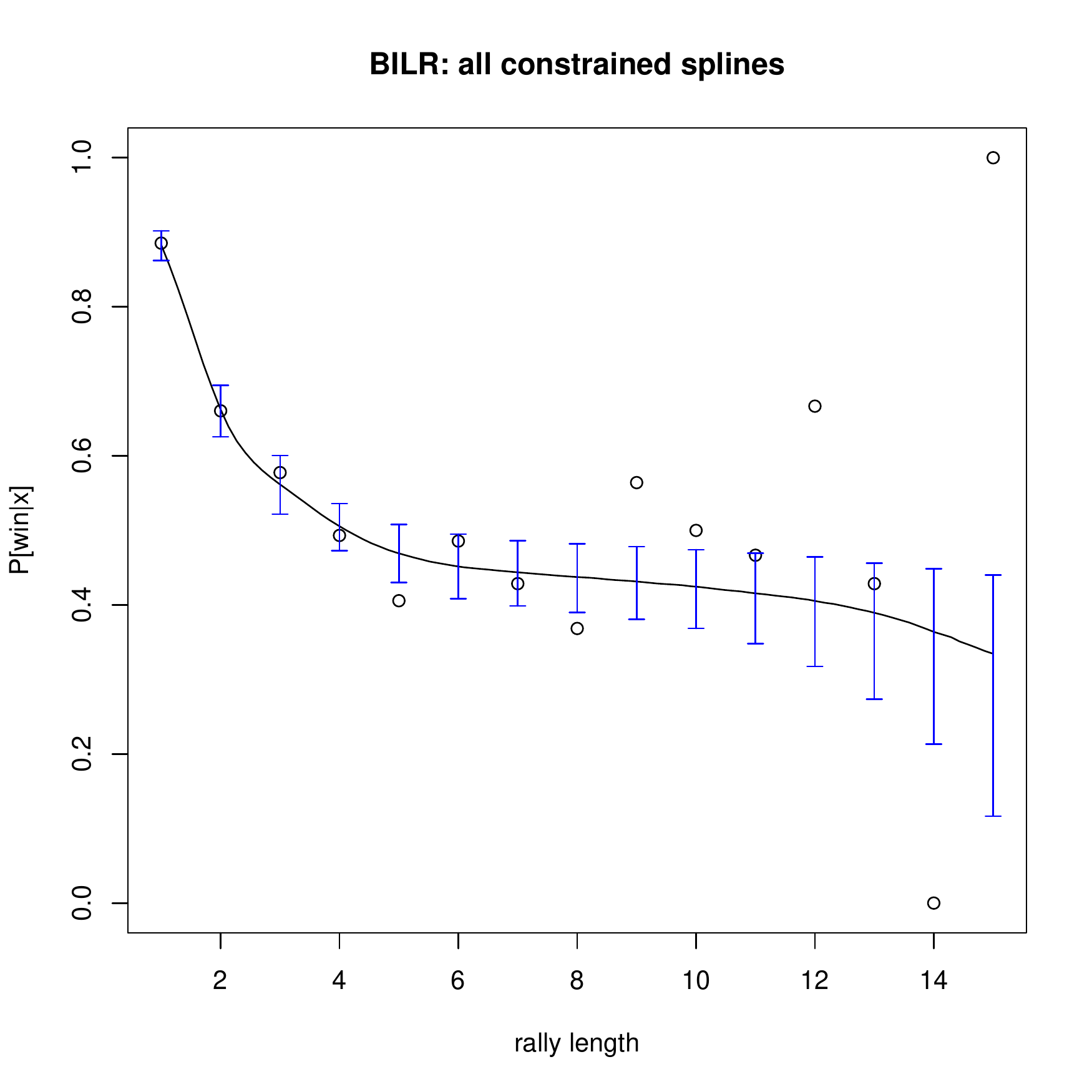}\\
		\end{center}
	\end{minipage}
	\caption{Probability of winning a point as a function of rally length for Andy Murray estimated with the exponential decay model (top left), BILR with no constrained splines (top right), BILR with order constraint on the coefficients of the splines with support in $(3, 15]$ (bottom left), and BILR with order constraint on all spline coefficients (bottom right). The points represents the real data, while the black lines are the posterior mean estimate of the probability of winning a point as a function of rally length obtained with these models. The blue dashed lines are the $95\%$ credible intervals. \label{fig:constrained}}
\end{figure}

In Figure \ref{fig:constrained} we compare the fit obtained for Andy Murray with the exponential model and our three versions of our BILR model. The exponential decay model (top left) has a decreasing behaviour until $x=5$, and after that the chance of winning a point  is a constant given by the difference between Murray's rally ability and the average rally ability of his opponents, i.e. to plot the figure we substituted $\alpha_j$ in Eq. \eqref{eq:mainmodel} by $\bar \alpha_i$ obtained averaging all the $\alpha_j$ for $j\neq i$. Our BILR model with no constraints (top right) allows for an increasing behaviour in the probability of winning the point for some intermediate values of rally length, and this behaviour is unlikely to be justified in practice. While for small values of rally length the decreasing trend in serve advantage can be learned from the data, for large values of rally length this behaviour must be imposed through the model given data sparsity. We recall that short rallies, i.e. $x\leq 4$, constitute $90\%$ of the rallies in the dataset. In this data-rich part of the domain, no constraint is needed to adequately describe the data. Conversely, for long rallies, the decreasing behaviour imposed via the prior on the coefficients leads to a model which is not influenced by outliers. Both the model with monotonicity constraint in $(3, 15]$ (bottom left) and the model with all spline coefficients constrained to be non-increasing (bottom right) display a non-increasing behaviour for large values of rally length.

For a quantitative evaluation of the performance of the four approaches, we compute the goodness-of-fit measures for these models, reported in Table \ref{tab:4}.
\begin{table}[!h]\centering
	\begin{tabular}{@{}lrrrrr@{}}\toprule
		& \multicolumn{4}{c}{\textbf{Goodness-of-fit-Measures}} \\
		\cmidrule{1-5} 
		& \textbf{LPML} & \textbf{WAIC} &\textbf{DIC}  &\textbf{RMSE}\\ \midrule
		\textbf{Exponential model} & -52813.7 & 105821.3  & 105876 & 22.52\\
		\textbf{Spline models}\\
		No constraints & -52760.4 & 105853.9  & 105818  & 20.71\\
		Non-increasing in $(3,15]$      & -52739.1 & 105747.6  & 105828 & 19.32\\
		Non-increasing in $[1,15]$    & -52744.9 & 105790.4  & 105875 & 20.37\\
		\bottomrule
	\end{tabular}
	\caption{Predictive goodness-of-fit measures for different model settings.}\label{tab:4}
\end{table}
Although no dramatic difference in performance emerge, the BILR model under setting 2) above (spline function non-increasing in $(3, 15]$) simultaneously maximises the LPML criterion and minimises the WAIC, DIC, and RMSE, respectively. According to the results in Table \ref{tab:4}, we select the model with six constrained splines. Thus, our final model has a spline function for server $i$:

\begin{equation}
f_i(s) = \sum_{m=1}^{9} \beta_{i, m} b_m(s), \quad \text{where} \quad s \in [1,15],
\end{equation}
with six constrained splines, that is, $ \beta_{i,4} \geq \beta_{i,5} \geq \beta_{i,6} \geq \beta_{i,7} \geq \beta_{i,8} \geq \beta_{i,9}$ for all servers $i=1,\dots, n_s$. 

\section{Results} \label{results}

In this Section, we report results of the model fitted to point-by-point data for main-draw singles Grand Slam matches from 2012 forward, which were described in Section~\ref{data}. We divide both the male and female datasets into training and test sets. In both training sets we have $90$ randomly chosen servers, while the receivers are $140$ in the male training set and $139$ in the female training set. Conversely, in the male test set we have $50$ servers and $140$ receivers, whereas in the female test set we have $49$ servers and $139$ receivers. We fit model \eqref{eq:likelihood}-\eqref{modelalpha} separately on both male and female training sets, and perform predictions on the hold-out test sets. Our aim is to predict the conditional probability of winning a point for servers in the test sets by borrowing information from the training set results. \par
The posterior update of the model parameters was performed via Gibbs sampling, implemented by the \texttt{rjags} package \citep{rjags} in the \texttt{R} programming language \citep{R}. Posterior summaries were based on $20,000$ draws from the posteriors, with a burn-in of 1000 iterations and thinning every 20 iterations to reduce the autocorrelation in the posterior samples. Convergence of the Markov Chain ha been assessed by visual inspection ad using the \texttt{coda} package \citep{coda2006}. The sampler appeared to converge rapidly and mix efficiently.

Summaries of the serve advantage model parameters suggest a strong serve advantage. The probability of the server winning conditional on the point ending on serve, $\mathbb{E}(Y_{ij} = 1 \vert x = 1)$, is $0.83$ with $95\%$ credible interval (C.I.) $(0.75,0.94)$ for men and $0.69$ for women with $95\%$ C.I. $(0.55,0.83)$, respectively. When the serve advantage is lost, e.g. $x=15$, the probability of winning a point is mainly given by the rally ability of the server against the rally ability of the opponent. In this case, $\mathbb{E}(Y_{ij} = 1 \vert x = 15)$ has credible intervals $(0.51,0.64)$ for men and $(0.46,0.55)$ for women, respectively. 
\begin{figure}[!h]
\centering
	\begin{minipage}{7cm}
		\begin{center}
			\includegraphics[width=7cm,height=6cm]{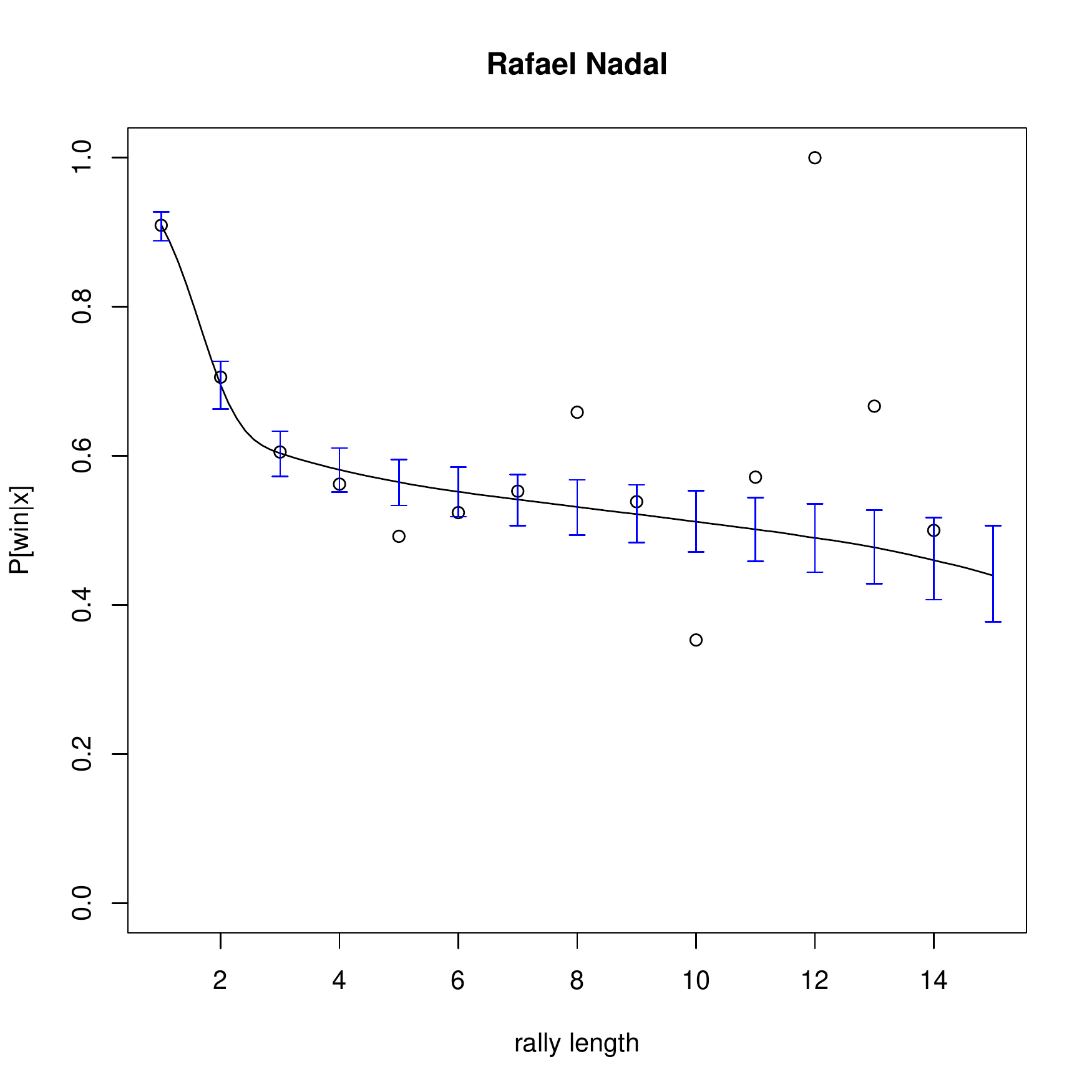}
		\end{center}
	\end{minipage}\ \
	\begin{minipage}{7cm}
		\begin{center}
			\includegraphics[width=7cm,height=6cm]{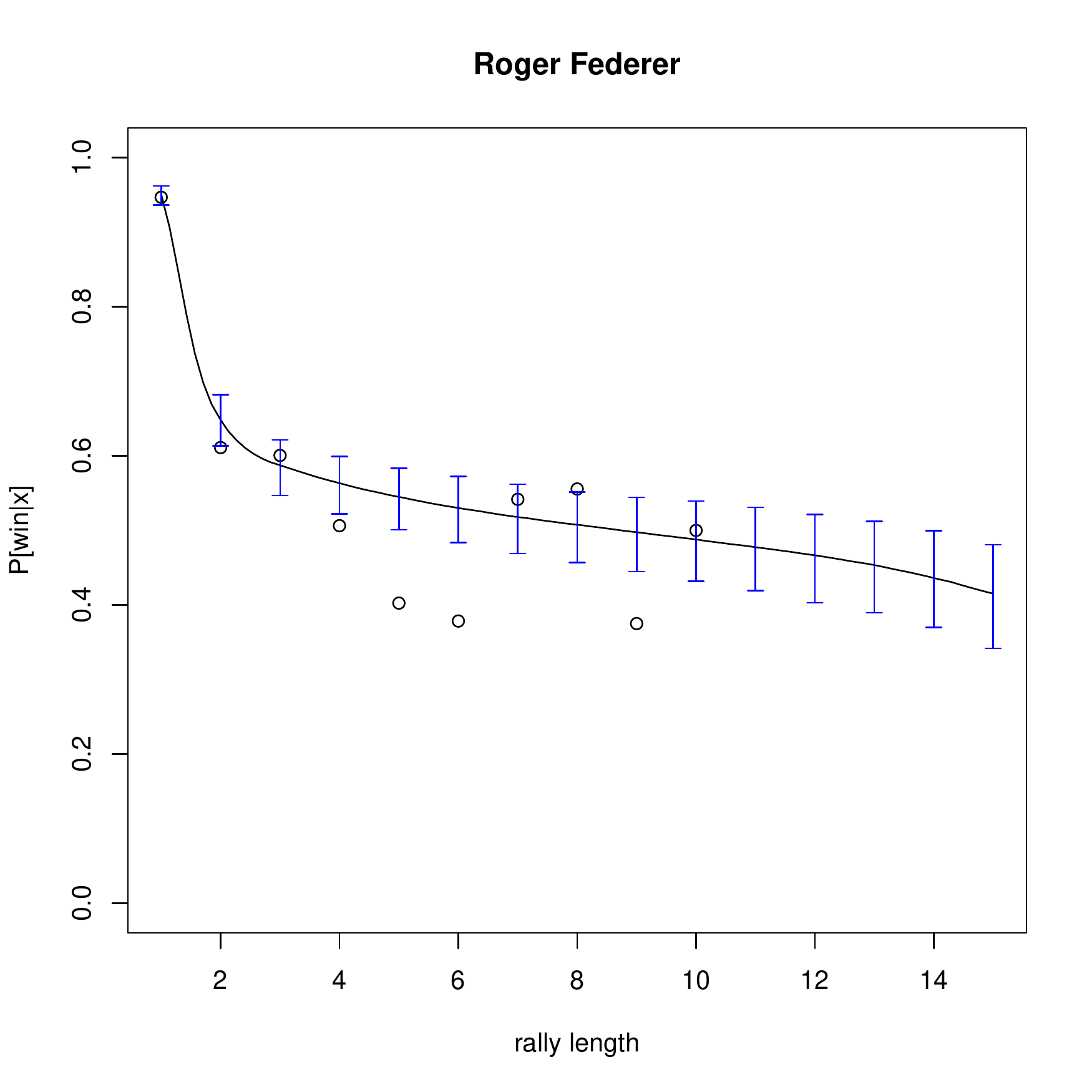}
		\end{center}
	\end{minipage}
	\caption{Probability of winning a point as a function of rally length for Rafael Nadal and Roger Federer. The points represents the real data, while the line is the posterior mean estimate of the probability of winning a point as a function of rally length obtained with the model. The blue dashed lines are the $95\%$ credible intervals. \label{fig:mod} }
\end{figure}
In Figure \ref{fig:mod} we show the estimated probability of winning a point as a function of rally length for Rafael Nadal and Roger Federer. The mean posterior curves are very similar: both players have the highest chance of winning the point on serve, and then this probability decreases. It is evident that the posterior mean estimate of the probability of winning a point on serve does not undergo an exponential decay, and a similar pattern for this estimate is observed on other players as well. % However, the curve which describes Roger Federer's play seems to follow a more pronounced decreasing behaviour compared to the Rafael Nadal's curve. Further, at large values of rally length the estimate is higher for Rafael Nadal than for Roger Federer, confirming the fact that the former is a good defender.  
We remark again that the curve $f_i(s)$ estimates a global trend, namely it describes how the serve advantage drops with rally length. Nevertheless, the value $f_i(x)$, for $x = 1,2, \dots, 15$, is the estimate of the serve advantage for athlete $i$ at the (discrete) value of rally length, $x$. In our Figure we decide to plot the posterior estimate of $f_i(s)$ as a continuous trajectory to underline the longitudinal structure of the data.

\begin{figure}[h!]
\centering
		\includegraphics[width=13cm,height=14cm]{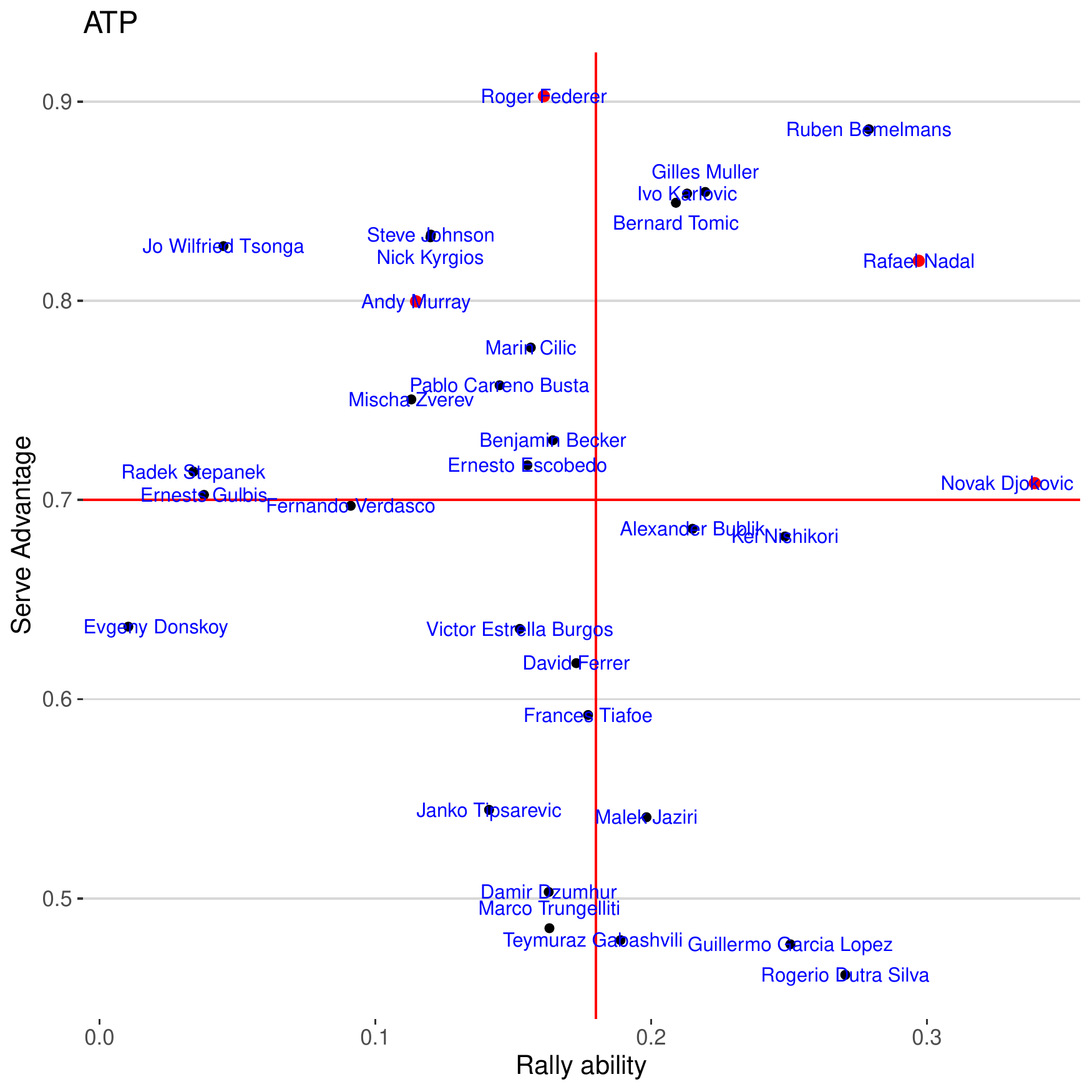}
	\caption{The posterior median rally ability $\alpha_i$ estimated under the baseline model (Section~\ref{sec:rallying}), on the $x$-axis, against the total serve advantage $\left(f_i(0) - f_i(15)\right)$ (Eq.~\eqref{eq:basis}), on the $y$-axis, for male players in the ATP tournaments. The red lines represent the median rally ability of male athletes, parallel to the $y$-axis, and the median of the serve advantage. The red points indicate the top four players of the ATP tournaments. We only display those athletes whose 95\% CI for serve advantage and rally ability do not include zero. \label{fig:ServevsRally} }
\end{figure}

\begin{figure}[h!]
	\centering
	\includegraphics[width=13cm,height=14cm]{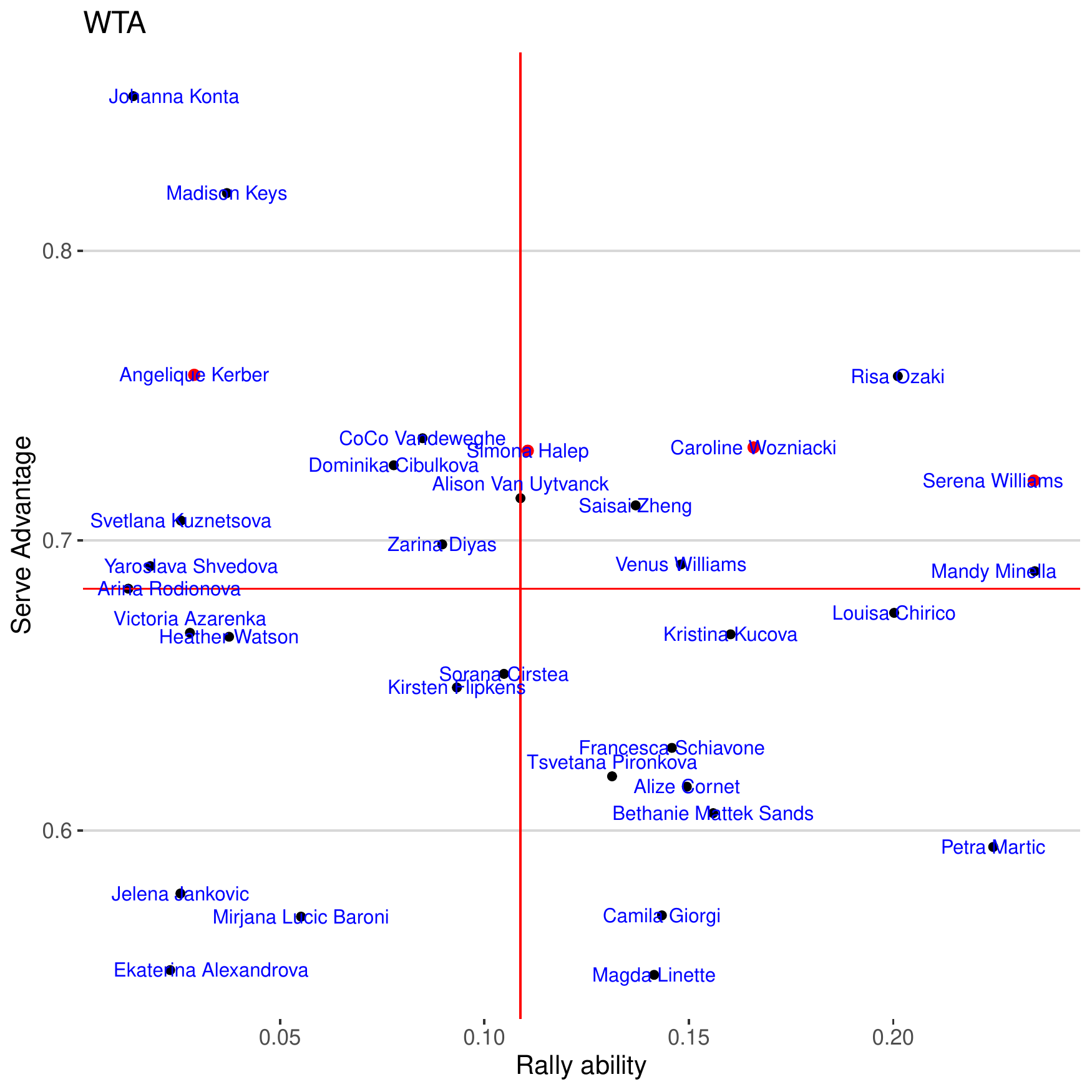}
	\caption{The posterior median rally ability $\alpha_i$ estimated under the baseline model (Section~\ref{sec:rallying}), on the $x$-axis, against the total serve advantage $\left(f_i(0) - f_i(15)\right)$ (Eq.~\eqref{eq:basis}), on the $y$-axis, for female players in the WTA tournaments. The red lines represent the median rally ability of female athletes, parallel to the $y$-axis, and the median of the serve advantage. The red points indicate the top four players of the WTA tournaments. We only display those athletes whose 95\% CI for serve advantage and rally ability do not include zero. \label{fig:ServevsRallyWTA} }
\end{figure}

Figure \ref{fig:ServevsRally} displays the estimated posterior median serve advantage versus the estimated posterior median rally ability. Specifically, the $x$-axis displays the posterior median of $\alpha_i$ estimated under the baseline model (Section~\ref{sec:rallying}), whereas the $y$-axis displays the total serve advantage $\left(f_i(0) - f_i(15)\right)$, where $f_i(s)$ is defined in Eq.~\eqref{eq:basis}. Let us observe the three top players according to the ATP singles ranking as of January 2019, namely, Roger Federer, Rafael Nadal and Novak Djokovic. We notice that Djokovic excels in terms of rally ability, confirming that he is better in defence than in attack. Conversely, Federer wins more at the first shot than on the long play. Finally, Nadal stands out in both terms of serve advantage and rally ability. Figure \ref{fig:ServevsRallyWTA} displays the same plot for the female dataset. We observe that Serena Williams excels on the long play, while Angelique Kerber and Simona Halep are better on serve. Caroline Wozniacki displays a good balance between serve and rally abilities.

Further, we want to investigate the effect of the surface on the rally ability. To this end we fit the extension of our model described in Section~\ref{sec:future}. Since the court is likely to have an effect on the player's rally abilities, we study how the court affects the players' skills. We report here the posterior median estimate for the rally ability along with 95\% credible intervals for the three different courts for the best players in the ATP and WTA tournaments, respectively. We also compute the posterior median estimate, with 95\% credible intervals, for ${\alpha_i}$, obtained with the model which does not take the court effect into account (Equation \eqref{eq:mainmodel}). These estimates, reported in Appendix~\ref{rallyabilities}, are used to rank the athletes and understand how the different courts impact to the game of these top players.

\begin{tiny}
	\begin{table}[h!]\centering
		\begin{tabular}{@{}lcllll@{}}\toprule
			& &  & \multicolumn{3}{c}{\textbf{Type of court}} \\
			\cmidrule{4-6} 
			& \multicolumn{1}{c}{\textbf{Ranking}} & \multicolumn{1}{c}{\textbf{Baseline}} & \multicolumn{1}{c}{\textbf{Clay}} & \multicolumn{1}{c}{\textbf{Grass}} & \multicolumn{1}{c}{\textbf{Hard}} \\ \midrule
			\textbf{ATP}\\
			& 1 & Djokovic (0.35)  & Nadal (0.52)    & Federer (0.28)  & Djokovic (0.34)\\
			& 2 & Nadal   (0.31)   & Djokovic (0.29) & Djokovic (0.28) & Nadal (0.28)\\
			& 3 & Federer  (0.16)  & Federer (0.09)  & Nadal (0.17)    & Federer (0.14)\\
			\textbf{WTA}\\
			& 3 & Wozniacki (0.17) & Halep  (0.20)   & Kerber (0.16)    & Wozniacki (0.21)\\
			& 1 & Halep (0.11)     & Wozniacki (0.09)& Halep (0.11)     & Kerber (0.15)\\
			& 2 & Kerber (0.03)    & Kerber (0.01)   & Wozniacki (0.11) & Halep (0.11)\\
			\bottomrule
		\end{tabular}
		\caption{Ranking of the players on different court surfaces. The second column lists the official ATP and WTA year-end final rankings (by points) for singles for the 2018 championships season.}\label{tab:6}
	\end{table}
\end{tiny}

The results (Table \ref{tab:6}) confirm common knowledge about these athletes. Djokovic and Nadal are both great at rallying. Djokovic is good on all courts, while Nadal is very good on clay and hard courts, but less favorite on grass. Federer appears to be weaker in rallying compared to the other two athletes, though he is the strongest on grass courts.
Regarding the WTA tournament, Angelique Kerber is good at rallying on both hard and grass court, but underperforming on clay courts. Caroline Wozniacki is good on all types of court, and in fact she is the player with the highest estimated rally ability $\alpha$ among the three female athletes. Simona Halep is good on clay, but does not outperform other players either on grass and hard courts. In general, however, the female athlete with the highest estimated $\alpha$ in the WTA dataset is Serena Williams (Figure~\ref{fig:ServevsRallyWTA}). The median estimate of her rally ability $\alpha$ in the baseline model is 0.26 (0.18-0.35), whereas the estimates on clay, grass, and hard courts are, respectively, 0.17 (0.05-0.30), 0.17 (0.11-0.21) and 0.27 (0.17-0.37).

\begin{figure}[!h]
	\centering
	\begin{minipage}{7cm}
		\begin{center}
			\includegraphics[width=7cm,height=6cm]{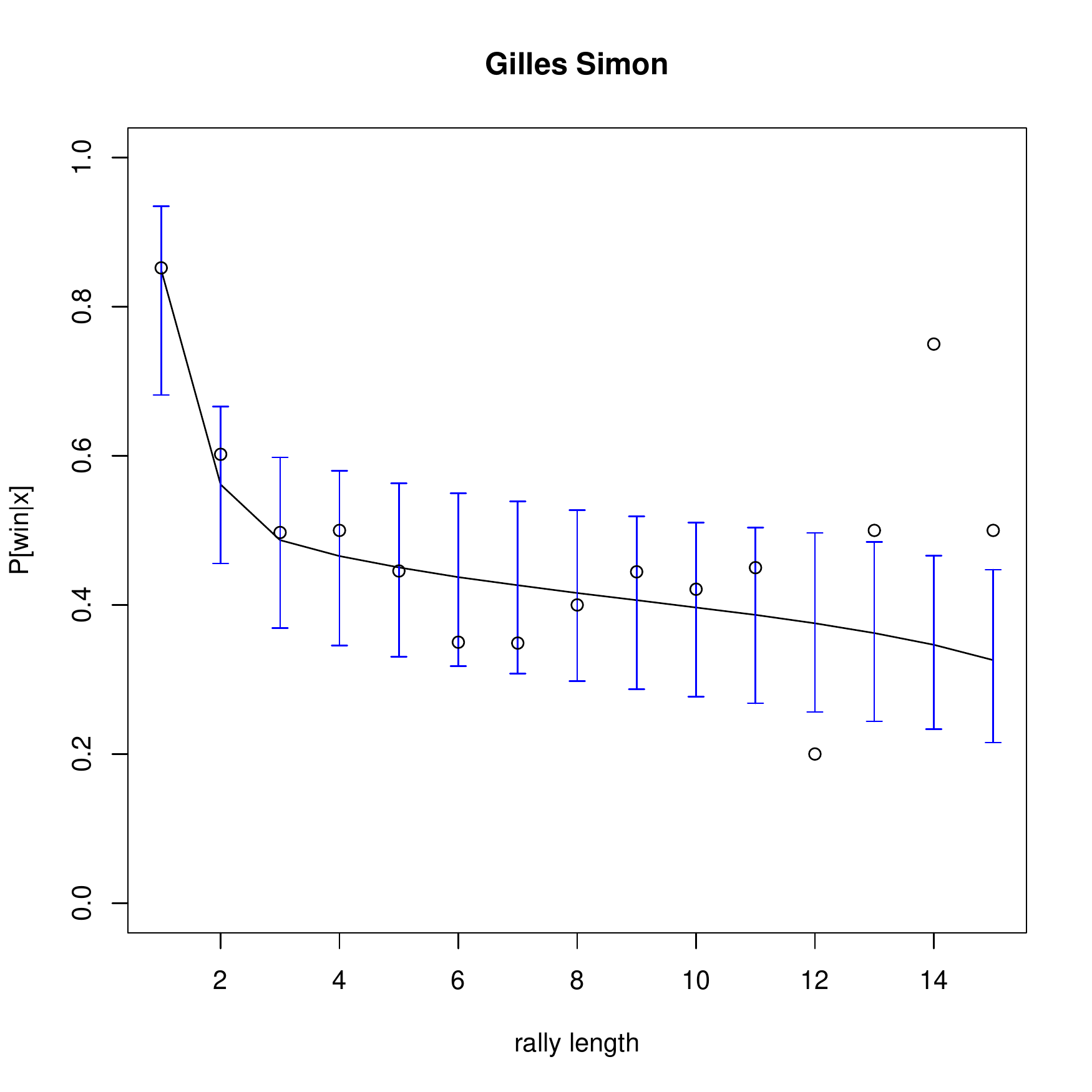}
		\end{center}
	\end{minipage}\ \
	\begin{minipage}{7cm}
		\begin{center}
			\includegraphics[width=7cm,height=6cm]{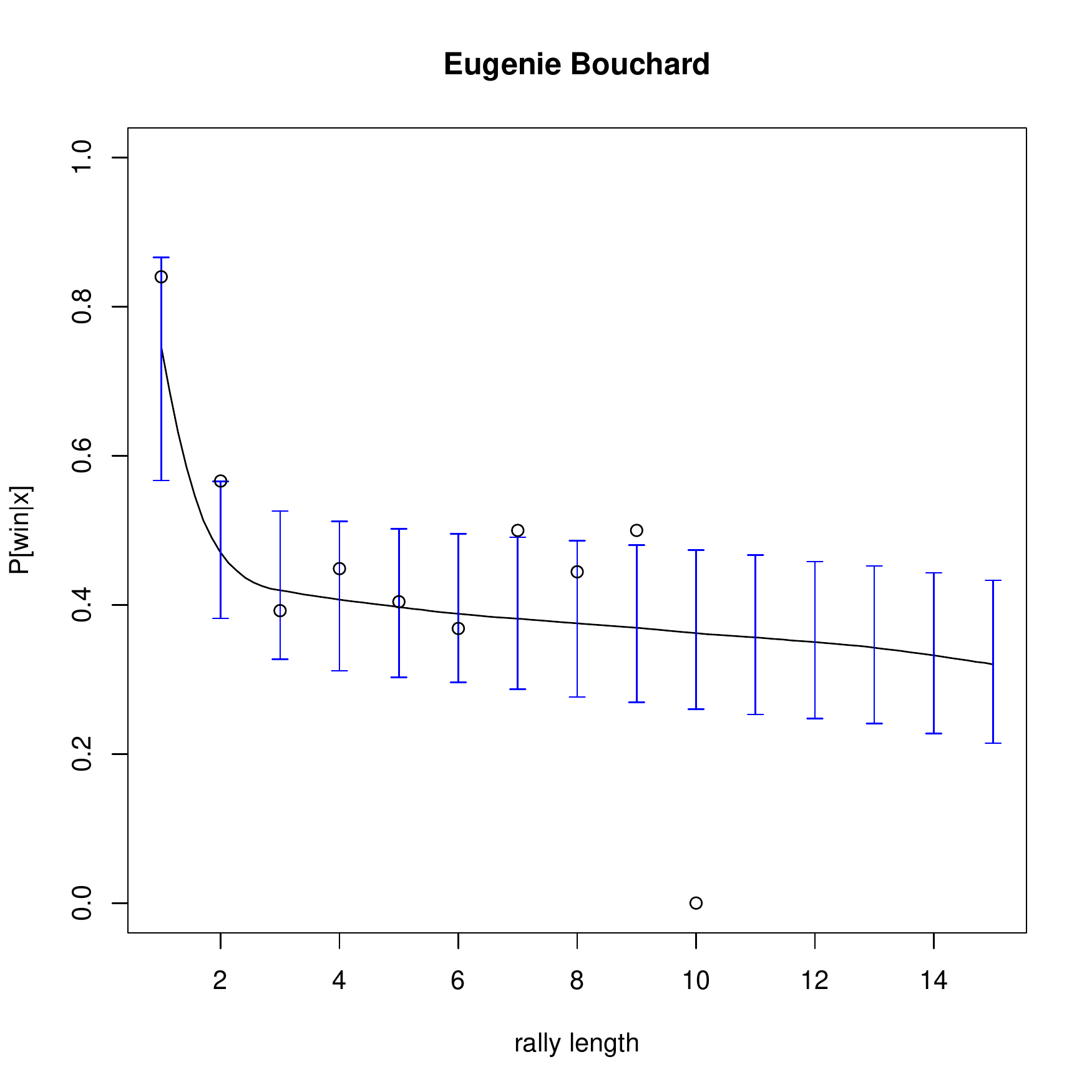}
		\end{center}
	\end{minipage}
	\caption{Probability of winning a point as a function of rally length for Gilles Simon, on the left, and for Eugenie Bouchard, on the right. The points represent the real data, while the line is the estimated posterior mean probability of winning a point as a function of rally length obtained with the model. The blue dashed lines are the $95\%$ credible intervals.} \label{fig:SimHal}
\end{figure}

In Figure \ref{fig:SimHal} we observe the out-of-sample prediction for two players belonging to the male and female test sets, Gilles Simon and Eugenie Bouchard. The estimated probability of winning a point for a server in the test set (e.g., the black solid curve in Figure \ref{fig:SimHal}), is obtained by drawing the basis functions coefficients $\beta_{i,1}, \dots, \beta_{i,M}$ and the positive random decrements $\{\varepsilon_{i,m}\}_{m = m_{L_0}+1}^{M}$ as per Equations \eqref{eq:priorbeta}, \eqref{eq:priorepsilon}, \eqref{eq:priorbeta2}, using the posterior estimates of the non-subject specific parameters, that is, $\beta_{m}$, $\sigma_{\beta_m}^2$, $r_{\varepsilon}$ and $s_{\varepsilon}$. The rally ability is just computed in the training phase. The strength of the hierarchical model is the ability to infer the conditional probability of winning for a hold-out subject by borrowing strength from athletes in the training dataset. The estimated trajectory for these players is in line with the observed realisations given by the points in Figure \ref{fig:SimHal}. \par

\section{Conclusions} \label{conclusions}

In this paper, we presented a framework to modelling the serve advantage in elite tennis. Our approach extends \cite{KOV} by replacing a simple decay exponential function for the serve advantage with a B-spline basis function decomposition, thus achieving more flexible results. Constraints on the basis function coefficients guarantee that the serve advantage is non-increasing with rally length. As in \cite{KOV}, we allow the conditional probability of winning on serve to also depend on the rally ability of the two players, and investigate how the different types of court may impact on such rally ability. When the exponential decay function in \cite{KOV} goes to zero, the conditional probability of winning a point is only given by the difference between two rally abilities, thus a constant. Conversely, our spline function is defined on $[1,15]$ by construction, and therefore non-zero everywhere in the spline domain. This results in higher uncertainty as represented by wider credible intervals for large values of rally length. % (see, for example, Figure \ref{fig:Nad}). 
This should be considered as a positive feature of our model, that is able to reflect larger uncertainty in presence of sparser data.

%\begin{figure}[!h]
%	\centering
%	\begin{minipage}{7cm}
%		\begin{center}
%			\includegraphics[width=7cm,height=6cm]{NadalCIK.pdf}
%		\end{center}
%	\end{minipage}\ \
%	\begin{minipage}{7cm}
%		\begin{center}
%			\includegraphics[width=7cm,height=6cm]{NadalCI.pdf}
%		\end{center}
%	\end{minipage}
%	\caption{Probability of winning a point as a function of rally length for Rafael Nadal. Estimates obtained with \cite{KOV}, on the left, and with our model, on the right. The points represent the real data, while the line is the estimated posterior mean probability of winning a point as a function of rally length. The blue segments are the point-wise $95\%$ credible intervals for each possible outcome of $x$. \label{fig:Nad}}
%\end{figure}

Our results show a sort of trade-off between serve advantage and rally ability. The most successful tennis players in the dataset show higher rally ability (rally ability above training median value) relative to their serve advantage. Indeed, if two players have the same chance of winning the point on the first shot, the match will be won by the player with the higher rally ability. We can conclude that although the service is important, what makes a tennis player great is his/her rally ability.\par
Although motivated by the analysis of tennis data, our methodology can be applied to pair-comparison data in general, with applications ranging from experimental psychology to the analysis of sports tournaments to genetics.

\appendix

  \section[]{Proof of Proposition \ref{prop:1}}\label{A}

Consider the restricted spline function 
\begin{equation*}
  f_{L_0,U}(s)= \sum_{m=m_{L_0}-k}^{M}\beta_m b_{m,k}(s),
\end{equation*}
where $k$ is the order of the B-splines. Following Formula (12) on page 116 and Formula (13) of \cite{deBoor}, we can compute its derivative as
\begin{align*}
  f'_{L_0,U}(s)&= (k-1)\sum_{m=m_{L_0}-k+1}^{M}\frac{\beta_m-\beta_{m-1}}{t_{m+k-1}-t_m} b_{m,k-1}(s)\\
  &\le (k-1) \sup_{m\in\{m_{L_0}-k+1,\dots,M\}}\left\{ 
    \frac{\beta_m-\beta_{m-1}}{t_{m+k-1}-t_m} 
  \right\} \sum_{m=m_{L_0}-k+1}^{M} b_{m,k-1}(s)\\
  &= (k-1) \sup_{m\in\{m_{L_0}-k+1,\dots,M\}}\left\{ 
    \frac{\beta_m-\beta_{m-1}}{t_{m+k-1}-t_m} 
  \right\} \\
\end{align*}
where we used the fact that that $\sum_{m=m_{L_0}-k+1}^{M} b_{m,k-1}(s)=1$. The latter follows from Formula (37) on page 96 of \cite{deBoor}, and from the fact that $b_{m,k-1}(s)=0$ for $s\in[L_0,U]$ and $m\notin\{m=m_{L_0}-k+1,\dots,M\}$.
%\citep[see Formula (37) page 96 of ][]{deBoor} for $s\in[L_0,U]$. \RAtext{Non so se serve dirlo: nella formula di de Boor gli indici vanno da 1 a $M$, ovvero sono tutte le splines. Ma per $s$ in $[L_0,U]$ le spline che non sono nella sommatoria qui, valgono zero}.
It is straightforward to observe that the constant $\sup_{m\in\{m_{L_0}-k+1,\dots,M\}}\left\{ 
    \frac{\beta_m-\beta_{m-1}}{t_{m+k-1}-t_m} 
  \right\} $ is smaller or equal than zero if and only if 
  \begin{equation*}
    \beta_m\le \beta_{m+1}\quad \text{for each}\quad m\in\{m_{L_0}+k-1\}
  \end{equation*}
The latter property is equivalent to requiring that the restricted control polygon $\mathcal{C}_{[L_0,U]}(s)$ is not increasing on its support, i.e. for $s\in[\bar{t}_{m_{L_0}-k},\bar{t}_M]$.

\section[]{Rally abilities on different types of courts }\label{rallyabilities}

\begin{table}[!h]\centering
	\begin{tabular}{@{}lrrrr@{}}\toprule
		&  & \multicolumn{3}{c}{\textbf{Type of court}} \\
		\cmidrule{3-5} 
		\textbf{Players}& \multicolumn{1}{c}{\textbf{Baseline}} & \multicolumn{1}{c}{\textbf{Clay}} & \multicolumn{1}{c}{\textbf{Grass}} &\multicolumn{1}{c}{\textbf{Hard}} \\ \midrule
		& \multicolumn{1}{c}{$\boldsymbol{\alpha}$} & \multicolumn{1}{c}{$\boldsymbol{\alpha_1}$} & \multicolumn{1}{c}{$\boldsymbol{\alpha_2}$}
		& \multicolumn{1}{c}{$\boldsymbol{\alpha_3}$}\\
		%\SIcanc{Andy Murray}     & 0.11 (0.03-0.29) & 0.14 (0.06-0.23) & 0.25 (0.17-0.33) & 0.06 (0.03-0.15) \\
		Novak Djokovic  & 0.35 (0.23-0.46) & 0.26 (0.17-0.36) & 0.25 (0.14-0.36) & 0.30 (0.22-0.40)\\
		Rafael Nadal    & 0.31 (0.19-0.40) & 0.49 (0.39-0.60) & 0.15 (0.03-0.28) & 0.25 (0.16-0.32)\\
		Roger Federer   & 0.16 (0.07-0.25) & 0.09 (0.01-0.21) & 0.27 (0.18-0.35) & 0.11 (0.03-0.20)\\
		Caroline Wozniacki& 0.17 (0.03-0.29) & 0.03 (0.00-0.11) & 0.11 (0.04-0.26) & 0.21 (0.11-0.31)\\
		Simona Halep    & 0.11 (0.03-0.21) & 0.20 (0.09-0.30) & 0.11 (0.01-0.22) & 0.11 (0.02-0.19)\\
		Angelique Kerber& 0.03 (0.01-0.10) & 0.09 (0.03-0.13) & 0.16 (0.06-0.32) & 0.15 (0.06-0.24) \\
		%Serena Williams & 0.26 (0.18-0.35) & 0.17 (0.05-0.30) & 0.17 (0.11-0.21) & 0.27 (0.17-0.37)\\
		%Maria Sharapova & 0.19 (0.09-0.25) & 0.04 (0.01-0.22) & 0.05 (0.01-0.21) & 0.13 (0.07-0.20)\\
		%Petra Kvitova   & 0.11 (0.02-0.19) & 0.03 (0.02-0.13) & 0.04 (0.02-0.14) & 0.18 (0.08-0.28)\\
		\bottomrule
	\end{tabular}
	\caption{Credible intervals for the rally abilities of the top players for the ATP and the WTA tournaments.}\label{tab:8}
\end{table}

\bibliographystyle{apa}
\bibliography{literature}

\end{document}